\documentclass[aps,rmp,10pt,twocolumn,floats,preprintnumbers,float fix,superscriptaddress,showpacs,showkeys,a4paper,twoside,longbibliography]{revtex4-1}
  
\usepackage{amsmath}
\usepackage{amssymb}
\usepackage{amsfonts}

\usepackage{time}

\newcommand{\threebox}[1]{\parbox{0.7cm}{{\tiny\ \hfill #1}}}
\newcommand{\fourbox}[1]{\parbox{2cm}{{\tiny\ \hfill #1}}}
\newcommand{\fivebox}[1]{\parbox{1cm}{{\tiny\ \hfill #1}}}
\newcommand{\largebox}[1]{\parbox{1cm}{\hbox{\footnotesize\ \hfill #1}}}

\usepackage{tikz}
\usetikzlibrary{decorations.pathmorphing}

\usepackage{times}
\usepackage{dcolumn}

\usepackage[english]{babel}

\usepackage{hyperref}

\def\ie{\emph{i.e.}\@~}
\def\cf{\emph{cf.}\@~}

\def\fig#1{\ref{fig:#1}}
\def\Fig#1{Fig.\@~\fig{#1}}

\def\Tab#1{Table~\ref{tab:#1}}

\def\eq#1{(\ref{eq:#1})}
\def\Eq#1{Eq.\@~\eq{#1}}
\def\Eqs#1{Eqs.\@~\eq{#1}}

\def\sec#1{\ref{sec:#1}}
\def\Sec#1{Sec.\@~\sec{#1}}

\def\CDF{\ensuremath{\mathcal{C}}}
\def\Rmin{\ensuremath{R_\text{min}}}
\def\Rmax{\ensuremath{R_\text{max}}}
\def\Rav{\ensuremath{\langle R \rangle}}

\def\Average#1{\ensuremath{\left\langle #1 \right\rangle}}

\def\rmd{\ensuremath{\mathrm{d}}}
\def\diffT{\ensuremath{\frac{\mathrm{d}}{\mathrm{d}t}}}

\def\mysection#1{\section{#1}\markright{\thesection . \ #1}}

\begin{document}
\title[Ripening and Focusing of Aggregate Size Distributions]{Ripening and Focusing of Aggregate Size Distributions\\ with Overall Volume Growth}

\author{J. Vollmer}
\affiliation{\text{Max Planck Institute for Dynamics and Self-Organization (MPI DS), G\"ottingen, Germany}}
\affiliation{\text{Faculty of Physics, Georg-August Univ.~G\"ottingen, G\"ottingen, Germany}}

\author{A. Papke}
\affiliation{\text{Max Planck Institute for Dynamics and Self-Organization (MPI DS), G\"ottingen, Germany}}
\affiliation{\text{Faculty of Physics, Georg-August Univ.~G\"ottingen, G\"ottingen, Germany}} 

\author{M. Rohloff}
\affiliation{\text{Max Planck Institute for Dynamics and Self-Organization (MPI DS), G\"ottingen, Germany}}
\affiliation{\text{Faculty of Physics, Georg-August Univ.~G\"ottingen, G\"ottingen, Germany}}

\begin{abstract}
  We explore the evolution of the aggregate size distribution in
  systems where aggregates grow by diffusive accretion of mass.
  Supersaturation is controlled in such a way that the overall
  aggregate volume grows linearly in time. Classical Ostwald ripening,
  which is recovered in the limit of vanishing overall growth,
  constitutes an unstable solution of the dynamics. In the
    presence of overall growth evaporation of aggregates always
  drives the dynamics into a new, qualitatively different growth
  regime where ripening ceases, and growth proceeds at a constant
  number density of aggregates. We provide a comprehensive description
  of the evolution of the aggregate size distribution in the constant
  density regime: the size distribution does not approach a universal
  shape, and even for moderate overall growth rates the standard
  deviation of the aggregate radius decays monotonically.  The
  implications of this theory for the focusing of aggregate size
  distributions are discussed for a range of different settings
  including the growth of tiny rain droplets in clouds, as long as
  they do not yet feel gravity, and the synthesis of nano-particles
  and quantum dots.
\end{abstract}


\date{\today\thinspace{}---\thinspace{}\now}

\maketitle
\markboth{J.~Vollmer, A.~Papke and M.~Rohloff: \ Ripening and Focusing of Aggregate Size Distributions}{}

\mysection{Introduction}
\label{sec:intro}

Characterising the evolution of the number density and the size
distribution of an assembly of aggregates in a fluid or solid matrix
has intrigued chemists
\cite{Wagner1961,Kahlweit1976b,JohnsonKorinekDongVeggel2012},
physicists \cite{LifshitzSlyozov1961,Bray1994,Slezov2009,Shneidman2013}, and
applied mathematicians
\cite{Voorhees1985,Penrose1997,NiethammerPego1999,Smereka2008,GoudonLagoutiereTine2012}
since it was first described by \citet{Ostwald1900}.
Early successes in the theoretical modeling focused on describing the diffusive
transport of material to the aggregates \cite{LaMerDinegar1950}.  In
many applications the volume fraction of the aggregates grows 
in time\thinspace{}---\thinspace{}either due to feeding by a chemical reaction, or
because temperature or pressure changes lead to a change of the
equilibrium volume fraction of the aggregates.  \citet{Reiss1951}
pointed out that the resulting sustained growth of the volume fraction
of the aggregates can lead to focusing of the aggregate size
distribution \citep[see][for recent
discussions]{KwonHyeon2011,ClarkKumarOwenChan2011,SowersSwartzKrauss2013}.
Subsequent theoretical work focused on the ripening of the aggregate
size distribution under thermodynamic equilibrium conditions, where to
a good approximation the aggregate volume fraction is preserved \cite{LifshitzSlyozov1961,Wagner1961}. This
dynamics involves aggregate ripening, a delicate balance of the evaporation of small
aggregates, and the redistribution of their volume to achieve further growth
of large aggregates. Assembly expectation values do not only change
due to the evolution of the shape of the size distribution, but also
by the change of its normalisation, i.e., the number of aggregates.
Independently, \citet{LifshitzSlyozov1961} and \citet{Wagner1961} derived scaling
laws for the decay of the number of aggregates, and the resulting growth
speed of the mean aggregate radius, and they determined the shape of the
asymptotic size distribution. Modern expositions derive their results
from the point of view of dynamic scaling theory
\cite{Voorhees1985,Bray1994,BarenblattBook}.

Here, we revisit the problem of simultaneous growth and coarsening
in the presence of overall volume growth. The increase of the aggregate volume
fraction  can be provided by different mechanisms:
\textbf{(i)}~a change of ambient temperature or pressure that drives
the system deeper into a miscibility gap \cite{vollmer97JCP2,cates03PhilTrans,vollmer07PRL},
\textbf{(ii)}~evaporation of small particles denoted
as \emph{sacrificial nano-particles}, that are continuously
added to the system \cite{JohnsonKorinekDongVeggel2012}, or 
\textbf{(iii)}~a chemical reaction or external flux of material into
the system \cite[\cf the review of][]{SowersSwartzKrauss2013}.
Depending on context the aggregates may be bubbles, droplets or solid aggregates.
However, in any case we consider aggregate growth for dilute systems
where merging of aggregates and sedimentation play a negligible
role.

We idealise aggregate growth and ripening by
considering the setting of a sustained constant flux onto the
aggregates \cite{NozawaDelvilleUshikiPanizzaDelville2005} which gives
rise to a linear growth of the aggregate volume fraction. For the
phase separation of binary mixtures such a setting has been studied
experimentally by \citet{auernhammer05JCP} and \citet{LappRohloffVollmerHof2012}.
The present work establishes that the net volume growth leads to
a cross over to behaviour that is remarkably different from the
behaviour assumed in dynamic scaling theory.

We present a new numerical algorithm that allows us to follow 
the aggregate growth up to five orders of magnitude in the volume --
\ie we cover a factor of $50$ in their average radius,~${\Rav}$.
This large range is needed to settle in the asymptotic scaling regime
where the form of the aggregate size distribution, and the exponents of
the power-law growth describing the aggregate number density and the
average volume can credibly be tested.
To gain insight into the impact of the net aggregate growth, we explore the
evolution of the size distribution for growth speeds, $\xi$, of the
aggregate volume fraction that cover a range of three orders of
magnitude.

\begin{figure}
{\footnotesize
  \input{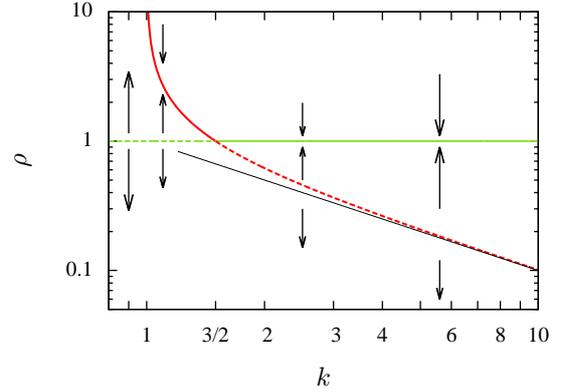}
}
\caption[]{Phase portrait of the evolution of the reduced aggregate radius $\rho = R/\Rav$.
  Dashed lines denote unstable fixed points, and solid line stable ones.
  The green lines denotes a fixed point at $\rho=1$, and the red lines the
  position of another fixed point, $\rho_+$, defined in \Eq{rho-pm}.  A thin straight black line has
  been added to show that $\rho_+$ rapidly approaches $k^{-1}$ for
  $k\gtrsim 5$. 
  \label{fig:phase_flow}}
\end{figure}

Based on our numerical study we set up a theoretical analysis that is
based on the evolution of the reduced aggregate radius, 
$\rho = R/{\Rav}$.  
In line with \citet{ClarkKumarOwenChan2011}'s findings the ratio 
\begin{equation}
  k = \frac{\Rav}{R_c} = 1 + \frac{\xi}{ 4\pi\; \sigma D \; n }
  \label{eq:define-k}
\end{equation}
of the average aggregate radius $\Rav$ and the critical
radius $R_c$, that separates the size of aggregates that grow from those
that shrink, is identified as the relevant control parameter that
governs the evolution.
For equilibrium systems the overall aggregate volume is preserved such
that $\xi=0$ and $k=1$. When there is a net growth of the overall
aggregate volume, the control parameter $k$ is increased by the ratio
of the growth rate $\xi$ and the diffusive relaxation rate of
supersaturation, $4\pi\, \sigma D \, n$ where $n$ is the number
density of aggregates, $D$ is the diffusion coefficient relevant for
the transport of material to the aggregates, and $\sigma$ is a length
scale of the order of the interface width \citep[\cf][and
\Sec{drop-growth} for details]{Bray1994,landau10}.
In \Fig{phase_flow} we provide a central result of the present study,
the phase portrait of the flow of $\rho$ at a constant $k$, which will
be derived and discussed in full detail in \Sec{reduced-size}.
Ripening at a fixed aggregate volume fraction, \ie for $\xi=0$,
amounts to the control parameter $k=1$. In this case $R_c = \Rav$ as
pointed out by \citet{LifshitzSlyozov1961}. For $\xi\simeq 0$, 
ripening arises by the interplay of an unstable fixed point of the
evolution for $\rho=1$ which enforces evaporation of small aggregates,
and the constraint of the overall conservation of volume that limits
the growth of the larger aggregates \cite[Chap.~7]{Slezov2009}.
Beyond $k=3/2$ this behaviour changes qualitatively due to an exchange
of stability bifurcation where the fixed point $\rho=1$ becomes
stable. In the following the consequences of this exchange on the
asymptotic form and evolution of the aggregate size distribution are
explicitly worked out, and compared to the numerical data.

The phase diagram, \Fig{phase_flow}, demonstrates how our
discussion provides a fresh view on a number of applications that
are under very active research presently:
A common feature of recipes for the synthesis of nano-particles
with narrow size distributions is that the focusing results from
aggregate growth proceeding in the presence of sustained mass flux,
that is reflected in an overall growth of the aggregate volume
\cite{NozawaDelvilleUshikiPanizzaDelville2005,ClarkKumarOwenChan2011,JohnsonKorinekDongVeggel2012,JanaSrivastavaPradhan2013}. 
In the chemical application one exploits transient focusing of the
polydispersity of the larger particles in bidisperse distributions
\cite{LudwigSchmelzer1985,JohnsonKorinekDongVeggel2012}, and in
systems where there is a considerable net flux onto the aggregates
\cite{Reiss1951,Sugimoto1987,PengWickhamAlivisatos1998,JanaSrivastavaPradhan2013}.
In these recipes the coarsening must be stopped once the chemical
precursor reaction that provides the material condensing on the
aggregates starts to cease.  We argue that this is done when $k$ drops
below $3/2$. Ripening would otherwise lead to a broadening of the very
sharp aggregate size distributions such that eventually they approach
the asymptotic \citet{LifshitzSlyozov1961} distribution 
\citep[see the review][]{SowersSwartzKrauss2013}.

Systems with a sustained flux onto the aggregates 
are also commonly encountered in the 
ripening and growth of bubbles in soda drinks, beer and
sparkling wine \cite{SoltzbergBowersHofstetter1997,ZhangXu2008}, and
in many natural processes.  Noticeable examples in the geo-sciences are the ripening and growth
of bubbles in the depths of geysers prior to eruption
\cite{IngebritsenRojstaczer1993,ToramaruMaeda2013,HanLuMcPhersonKeatingMooreEtAl2013},
and the growth of bubbles \cite{Manga1996} and crystallites in
cooling magma \cite{Sparks1987,MartinNokes1988}.

The paper is organised as follows: 
In \Sec{aggregate-size} we derive the equations of motion for the aggregate
radius, and explain how the equations are integrated numerically.
For $N$ aggregates the evolution is provided by a set of $N$
non-linear differential equations for the respective radii. The
equations are coupled because they involve moments of the size
distribution.  A theoretical description of the time evolution of the
aggregate size distribution is obtained in three steps: In
\Sec{average-size} we explore the time evolution of the relevant
moments of the aggregate size distribution. This allows us in
\Sec{reduced-size} to solve the evolution of the size of individual
aggregates \emph{constrained} to the time evolution of the
moments. Hence, we reduce the problem of solving the set of $N$
equations to finding the solutions of a single non-linear differential
equation for $N$ different initial conditions, which define the
initial aggregate size distribution. At this point we also explore the
consequences the exchange of stability bifurcation on the evaporation
of aggregates.
Subsequently, in \Sec{distribution} we combine the results on the
evolution of the moments and on the resulting evolution of the size of
individual aggregates to obtain the evolution of the aggregate size
distribution.
In each step of this analysis we compare the predictions to the
numerical data.
The implications of our findings on different experimental systems are
discussed in \Sec{discussion}, and the the prime results of our study
are summarised in \Sec{conclusion}.

\mysection{The assembly of aggregate radii}
\label{sec:aggregate-size}

In principle many different processes contribute to aggregate growth.
Here, we consider the case where 
\begin{itemize}
\item there are sufficiently few aggregates such that they grow by
  diffusive flux received from a mean-field background supersaturation
  field\thinspace{}---\thinspace{}analogously to Lifshitz-Slyozov-Wagner theory \cite{Bray1994}
\item the feeding rate, $\xi$, is sufficiently small such
  that it only affects the mean-field level of supersaturation, and 
  does not interfere with the diffusion coupling the aggregates to the
  supersaturation \cite[\cf][for a discussion of potential changes to the diffusion equation]{Vollmer2008}.
\end{itemize}

\subsection{Evolution of the aggregate radii and their volume}
\label{sec:drop-growth}

The supersaturation in the bulk is relaxed by diffusion onto the
aggregates, causing them to grow. 
Following \citet{landau10,Bray1994}, we have
\begin{equation}
  \dot R = \frac{\sigma D}{R^2} \; \left[ \frac{R}{R_c} -1 \right] \, .
  \label{eq:Rdot}
\end{equation}
Here, $R_c$ is the critical aggregate radius which depends on the
supersaturation in the system, $D$ is the pertinent concentration
diffusion coefficient, and $\sigma$ is a microscopic length scale
which accounts for the aggregate-size dependence of the chemical
potential drop that is driving the diffusive fluxes. Specifically, $\sigma$ is
proportional to the interfacial tension. Its full parameter
dependence and characteristic values for some typical applications are
provided in \Sec{discussion}.

The term in square brackets in \Eq{Rdot} accounts for the effect of
interfacial tension on aggregate growth.  Interfacial tension penalises
small aggregates such that only aggregates with a radius larger than
$R_c$ can grow.  For instance, in Lifshitz-Slyozov-Wagner theory no
supersaturation is provided externally, and
$R_c$ is equal to
the average radius $\Rav$.  Smaller aggregates evaporate, and hence
they provides the supersaturation which admits the growth of the
larger aggregates.

Let us now consider the evolution of $N$ aggregates of respective radius
$R_i$, $i=1\dots N$. Their total volume is
\begin{subequations}
\begin{equation}
   V = \frac{4\pi}{3} \sum_{i=1}^N R_i^3 \, .
\label{eq:SumRi}
\end{equation}
Introducing the average aggregate radius, ${\Rav} = N^{-1} \sum_i R_i$, 
one finds 
\begin{eqnarray}
\dot V &=& 4\pi \sum_i R_i^2 \dot R_i 
       = 4\pi\; \sigma D \; \sum_i \left[ \frac{R_i}{R_c} -1 \right]
\nonumber
\\[2mm]
       &=& 4\pi\; \sigma D \; N\;  \left[ \frac{{\Rav}}{R_c} -1 \right] 
        =  4\pi\; \sigma D \; N\; (k -1 ) \, ,
\label{eq:drop-vol-growth}
\end{eqnarray}%
\end{subequations}%
where we have used the definition $k = {\Rav}/{R_c}$ in the last step (\cf\Eq{define-k}).
Here and in the following the brackets \Average{.} denote the average over the
aggregate assembly, 
\[
\Average{ f(R) } := \frac{1}{N} \; \sum_i  f( {R}_i ) \, .
\]
In particular, ${\Rav}$ is the average aggregate radius,
and 
\begin{equation}
\Average{ R^3 } = \frac{3V}{4\pi\,N} =  3 \, \sigma D \; (k -1 ) \; t \,.
\label{eq:avRcube}
\end{equation}
There is no constant term in this equation due to an appropriate choice of 
the initial time $t_0$ such that the initial volume $V_0$ amounts to 
\[
V_0 = \frac{4\pi}{3} \sum_i  \left( R_i(t_0) \right)^3 = 4\pi\; \sigma D \; N_0\; (k -1 ) \, t_0 \, .
\]
The linear growth $\xi$ of the aggregate volume fraction $V/\mathcal{V}$ in a system 
of sample volume $\mathcal{V}$ amounts to
\begin{subequations}
\begin{equation}
  V = V_0 + \mathcal{V} \xi ( t - t_0 ) \,.
  \label{eq:lin-V-growth}
\end{equation}
Together with \Eq{drop-vol-growth} this growth implies, 
\begin{eqnarray}
  \mathcal{V} \xi &:=& \dot V = 4\pi\; \sigma D \; N\; (k-1)
  \label{eq:Vdot}
\end{eqnarray}%
\end{subequations}%
such that we derive here the dependence anticipated in \Eq{define-k}.

Altogether, we find the following set of equations for the
evolution of the aggregate radii, $R_i$,
\begin{eqnarray}
\dot R_i &=& 
\frac{\sigma D}{R_i^2} 
 \;  \left[ k \; \frac{R_i}{{\Rav}} -1 \right]\, , \quad i=1\dots N \, ,
\label{eq:Ri_evolution}
\end{eqnarray}
where $k$ is a function of the growth rate $\xi$, as stated in \Eq{define-k}.
The growth of the aggregate radii, $R_i$, is coupled in a mean-field way via the dependence of the
equations on the average aggregate radius ${\Rav}$, and via
$k$ also explicitly on the number, $N = n\mathcal{V}$, of aggregates.

\subsection{Numerical implementation}
\label{sec:Implementation}

\begin{figure}
\setlength{\unitlength}{2700sp}%
\begingroup\makeatletter\ifx\SetFigFont\undefined%
\gdef\SetFigFont#1#2#3#4#5{%
  \reset@font\fontsize{#1}{#2pt}%
  \fontfamily{#3}\fontseries{#4}\fontshape{#5}%
  \selectfont}%
\fi\endgroup%
\begin{picture}(5904,9024)(409,-8073)
\thicklines
{\color[rgb]{0,0,0}\put(1351,389){\framebox(1800,450){}}}%
{\color[rgb]{0,0,0}\put(1126,-511){\framebox(2250,450){}}}
{\color[rgb]{0,0,0}\put(1126,-1411){\framebox(2250,450){}}}
{\color[rgb]{0,0,0}\put(1126,-2311){\framebox(2250,450){}}}
{\color[rgb]{0,0,0}\put(4051,-2336){\framebox(2250,600){}}}
{\color[rgb]{0,0,0}\put(2251,-2761){\line(-4,-3){900}} 
                   \put(1351,-3436){\line( 4,-3){900}}
                   \put(2251,-4111){\line( 4, 3){900}}
                   \put(3151,-3436){\line(-4, 3){900}}}%
%
{\color[rgb]{0,0,0}\put(1076,-5236){\framebox(2400,675){}}}
{\color[rgb]{0,0,0}\put(2235,-5667){\line(-5,-4){877.561}} 
                   \put(1351,-6361){\line( 4,-3){880.640}}
                   \put(2235,-7017){\line( 4, 3){901.120}}
                   \put(3151,-6361){\line(-4, 3){919.360}}}%
%
{\color[rgb]{0,0,0}\put(1726,-7950){\framebox(1050,450){}}}
%
%
%
%
{\color[rgb]{0,0,0}\put(2251,-556){\vector( 0,-1){360}}}%
{\color[rgb]{0,0,0}\put(2251,344){\vector( 0,-1){360}}}%
{\color[rgb]{0,0,0}\put(2251,-1456){\vector( 0,-1){360}}}%
{\color[rgb]{0,0,0}\put(2251,-2356){\vector( 0,-1){360}}}%
{\color[rgb]{0,0,0}\put(4951,-1546){\line( 0, 1){385}}
                   \put(4951,-1161){\vector(-1, 0){1500}}}%
{\color[rgb]{0,0,0}\put(1261,-6361){\line(-1, 0){810}}
                   \put(451,-6361){\line( 0, 1){6075}}
                   \put(451,-286){\vector( 1, 0){580}}}%
{\color[rgb]{0,0,0}\put(2251,-4156){\vector( 0,-1){360}}}%
{\color[rgb]{0,0,0}\put(2251,-7081){\vector( 0,-1){360}}}%
{\color[rgb]{0,0,0}\put(2251,-5281){\vector( 0,-1){360}}}%
{\color[rgb]{0,0,0}\put(3376,-3436){\line( 1, 0){1575}}
                   \put(4951,-3436){\vector( 0, 1){855}}}%
\put(2341,-4336){\makebox(0,0)[lb]{\smash{{\SetFigFont{10}{12.0}{\familydefault}{\mddefault}{\updefault}{\color[rgb]{0,0,0}yes}%
}}}}
\put(2341,-7216){\makebox(0,0)[lb]{\smash{{\SetFigFont{10}{12.0}{\familydefault}{\mddefault}{\updefault}{\color[rgb]{0,0,0}no}%
}}}}
\put(3466,-3616){\makebox(0,0)[lb]{\smash{{\SetFigFont{10}{12.0}{\familydefault}{\mddefault}{\updefault}{\color[rgb]{0,0,0}no}%
}}}}
\put(4150,-1940){\makebox(0,0)[lb]{\smash{{\SetFigFont{10}{12.0}{\familydefault}{\mddefault}{\updefault}{\color[rgb]{0,0,0}evaporation:}%
}}}}
\put(4150,-2210){\makebox(0,0)[lb]{\smash{{\SetFigFont{10}{12.0}{\familydefault}{\mddefault}{\updefault}{\color[rgb]{0,0,0}update  $dV$,  $N$}%
}}}}
\put(2251,-331){\makebox(0,0)[b]{\smash{{\SetFigFont{10}{12.0}{\familydefault}{\mddefault}{\updefault}{\color[rgb]{0,0,0}calculate  $dV$, \Eq{Vdot}}%
}}}}
\put(2251,-1231){\makebox(0,0)[b]{\smash{{\SetFigFont{10}{12.0}{\familydefault}{\mddefault}{\updefault}{\color[rgb]{0,0,0}calculate  $k$, \Eq{define-k}}%
}}}}
\put(2251,-2131){\makebox(0,0)[b]{\smash{{\SetFigFont{10}{12.0}{\familydefault}{\mddefault}{\updefault}{\color[rgb]{0,0,0}set trial  $\{Q_i(t+dt)\}$}%
}}}}
\put(2251,-3230){\makebox(0,0)[b]{\smash{{\SetFigFont{10}{12.0}{\familydefault}{\mddefault}{\updefault}{\color[rgb]{0,0,0} all  }%
}}}}
\put(2251,-3510){\makebox(0,0)[b]{\smash{{\SetFigFont{10}{12.0}{\familydefault}{\mddefault}{\updefault}{\color[rgb]{0,0,0}$Q_i(t+dt)>0$}%
}}}}
\put(2251,-3790){\makebox(0,0)[b]{\smash{{\SetFigFont{10}{12.0}{\familydefault}{\mddefault}{\updefault}{\color[rgb]{0,0,0} ?    }%
}}}}
\put(1150,-4786){\makebox(0,0)[lb]{\smash{{\SetFigFont{10}{12.0}{\familydefault}{\mddefault}{\updefault}{\color[rgb]{0,0,0}accept trial}}}}}
\put(1150,-5056){\makebox(0,0)[lb]{\smash{{\SetFigFont{10}{12.0}{\familydefault}{\mddefault}{\updefault}{\color[rgb]{0,0,0}update $t$, $\{R_i:=Q_i^{1/3}\}$}}}}}
\put(2251,-6406){\makebox(0,0)[b]{\smash{{\SetFigFont{10}{12.0}{\familydefault}{\mddefault}{\updefault}{\color[rgb]{0,0,0}go on?}%
}}}}
\put(2251,-7780){\makebox(0,0)[b]{\smash{{\SetFigFont{10}{12.0}{\familydefault}{\mddefault}{\updefault}{\color[rgb]{0,0,0}end}%
}}}}
\put(2251,569){\makebox(0,0)[b]{\smash{{\SetFigFont{10}{12.0}{\familydefault}{\mddefault}{\updefault}{\color[rgb]{0,0,0} initialise, \Eq{IC}}%
}}}}
\put(1036,-6541){\makebox(0,0)[lb]{\smash{{\SetFigFont{10}{12.0}{\familydefault}{\mddefault}{\updefault}{\color[rgb]{0,0,0}yes}%
}}}}
\end{picture}%

\caption[]{Schematics of the integration scheme for
  the size distribution $\{ R_i \}_{i=1\dots N}$, where the aggregate
  number $N$, the volume increments $\rmd V$ and the parameter $k$ are self-consistently
  adjusted when small aggregates evaporate.
\label{fig:program_outline}}
\end{figure}
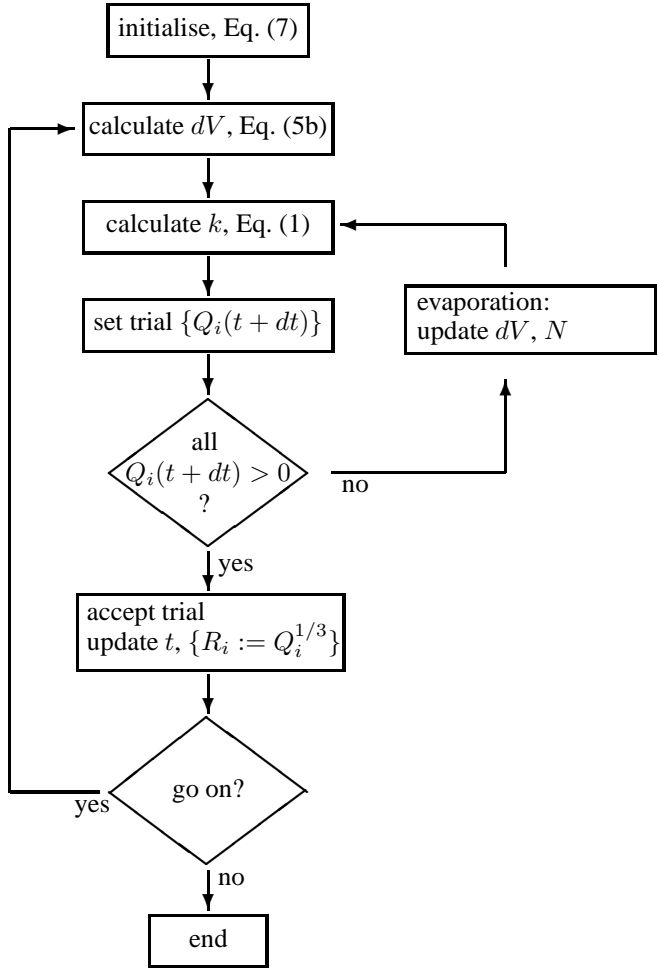

The implementation of the integration scheme is detailed in the flow
chart provided in \Fig{program_outline}.
To follow the size evolution of an assembly of aggregates, we integrate
the cubes, $Q_i := R_i^3$ of their respective radii. This avoids
instabilities in the numerics arising when directly integrating
\Eq{Ri_evolution} for very small aggregates. 
In each time step we calculate the radii, $R_i$ and their mean value,
${\Rav}$, and determine the updates of the $Q_i$ via a
predictor-corrector scheme that keeps track of the growth of the
overall aggregate volume, \Eq{drop-vol-growth}.  It uses a recursion
to identify and remove aggregates that evaporate in a given time step.
Prior to calculating ${\Rav}$ and using \Eq{Ri_evolution} to determine
the respective volume increments, the volume of evaporating aggregates
is transferred to the volume increment to be added to the surviving
aggregates.

\begin{figure*}
{\footnotesize
  \input{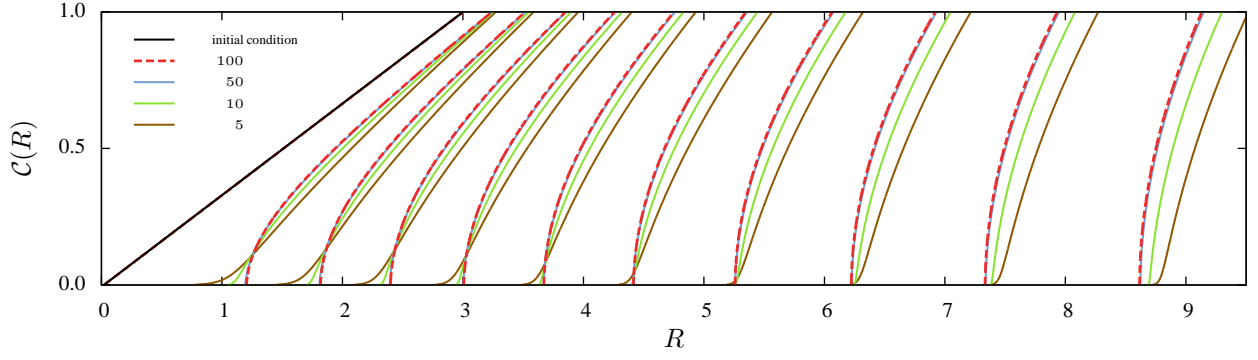}
}
\caption[]{The four sets of curves of different colour show
    stroboscopic snapshots of the time evolution of the cumulative
    size distribution function, $\mathcal{C}(R)$, of aggregates for
    the same initial condition, and $k=5$, $10$, $50$, and $100$,
    respectively. Here and in the following we use dashed lines for
    the largest value of $k$ displayed in the plot, and solid lines
    for all other curves. We use the same colour for all data
    referring to a given value of $k$, and provide the initial
    conditions, \Eq{IC}, by a solid black line (the leftmost curve).
    The time increments between successive curves of the same colour
    correspond to a time lapse resulting in an increase of the total
    aggregate volume by a factor of $10^{1/5}$.  Consequently, the
    rightmost curves of each colour correspond to systems where the
    total aggregate volume grew by a factor of hundred.  In the main
    text we discuss the similarities and differences between the CDFs
    in each of the resulting quadruplets. This allows us to
    pinpoint salient features of the impact of $k$ on the time
    evolution of the CDFs.
    \label{fig:CDFevolution}}
\end{figure*}

All numerical data in the present paper refer to an initial assembly of $N_0$
aggregates with a distribution that is flat in the radius between
$R = \Rmin \dots \Rmax$, 
\begin{subequations}
\begin{eqnarray}
  && R_i = \Rmin + (\Rmax - \Rmin) \; \frac{i - 1}{N_0 - 1} \, , 
  \;\; i = 1\dots N_0 
  \\[2mm]
  && \text{with} \quad  
  N_0 = 1000\, , \quad 
  \Rmin = 0.02\, , \quad
  \Rmax = 3.00 \, .
\end{eqnarray}%
\label{eq:IC}%
\end{subequations}%

We make use of the linear growth of the overall aggregate volume,
\Eq{lin-V-growth}, to specify the elapsed time in terms of the average
aggregate volume, and choose the scale for the aggregate radius such
that \hbox{$\sigma D \equiv 1$}.

For the bookkeeping of evaporation of aggregates we observe that the
increasing order of the aggregate radius with index $i$ is preserved by
the evolution. After all, \Eq{Ri_evolution} implies that
\begin{equation}
  Q_i > Q_j 
  \Rightarrow \diffT (Q_i - Q_j) 
  = \frac{3\sigma D \, k}{\Rav} \; (R_i - R_j) > 0
\label{eq:sortedR}
\end{equation}
such that the difference of the aggregate volumes grows strictly
monotonously. Consequently, the evaporation of aggregates can
conveniently be taken into account in our algorithm by appropriately
truncating the range of the index $i$.

The algorithm admits adaptive step size control.  After some testing
we decided however to rather choose equidistant time steps on a
logarithmic time axis because this saves the numerical overhead of the
adaptive step size control and is convenient for the data analysis.
For all data shown in this paper we took $10^6$ integration steps to
increase the aggregate volume by one order of magnitude. This provides
an accurate and very fast integration routine, where the simulation
can span many orders of magnitude of aggregate growth.

Figure~\fig{CDFevolution} shows the evolution of the cumulative
aggregate size distribution (CDF), $\mathcal{C}(R)$, for four
different values of $\xi$ that correspond to initial values of $k=5$,
$10$, $50$, and $100$.  The CDF provides the fraction of aggregates
with a radius smaller than $R$.
Hence, for the flat initial distribution, \Eq{IC}, the initial CDF
amounts to a function that rises linearly from zero at $\Rmin = 0.02$
to one at $\Rmax = 3.00$\,. This initial CDF is shown by the solid
black line at the smallest values of $R$.  To the right of this
initial condition we show ten quadruples of functions displaying the
respective CDFs at later times. Each set allows us to compare the
shape of the CDF in a situation where the overall volume of the
aggregates matches, \ie for the same dimensionless time in our
simulations.  At this point we make four observation that we will be
further substantiated in the forthcoming discussion:
\begin{itemize}
\item At early times the distributions for $k=5$ and $10$ develop a
  tail towards the small aggregates, and they feature larger average
  aggregate sizes at late times. This is a hallmark of the evaporation
  of aggregates. The tail is due to aggregates that shrink and
  evaporate when their radius approaches zero. The larger average size
  is required to achieve a prescribed overall volume with a smaller
  number of aggregates.
\item The CDFs for $k=50$ and $100$ look almost the same.  Indeed,
  this holds for all $k\gtrsim 50$, where no aggregates evaporate.
\item From the inspection of the numerical data one verifies that for
  all $k > 1$ the growth at late times proceeds at a fixed aggregate
  number. Subsequently, the difference in shape with respect to the
  CDFs for larger values of $k$ does not evolve any longer.
\item All distributions become more and more monodisperse.
\end{itemize}
The evolution of the size of individual aggregates and their
evaporation is discussed in \Sec{evaporation}, and in
\Sec{distribution} we address the time evolution of the CDFs.
These results rest upon a priori insights into the time evolution of
the moments of the aggregate size distribution that are supplied in
\Sec{average-size}.

\mysection{Moments of the aggregate size distribution}
\label{sec:average-size}

The set of differential equations \eq{Ri_evolution} can be decoupled
when the time evolution of $N$ and \Rav\ can be determined a priori,
\ie without explicitly integrating the set of equations $\dot R_i$.
Our numerics revealed that for all $k>1$ the number of aggregates $N$
is constant at late times, and that for sufficiently large $k$ there
is no evaporation at all.
In this section we therefore establish the time evolution of \Rav\ for
a constant number of aggregates,~$N$.

\subsection{Asymptotic evolution of ${\Rav}^2 \diffT{{\Rav}}$}

For a constant number of particles the time derivative of the average aggregate radius
\[
  {\Rav}  = \frac{1}{N} \; \sum_i R_i \, ,
\]
based on \Eq{Ri_evolution} is given by
\begin{subequations}
\begin{eqnarray}
  \diffT{{\Rav}}  
  &=&  \frac{1}{N} \; \sum_i  \dot{R}_i
  = \frac{1}{N} \; \sum_i \frac{\sigma D}{R_i^2} \;  \left[ k \; \frac{R_i}{{\Rav}} -1 \right]
  \\[2mm]
  &=& \frac{\sigma D}{{\Rav}^2} \;  
       \left[ k \; \Average{   R^{-1}  } \; {\Rav} 
              -    \Average{   R^{-2}  } \; {\Rav}^2 
       \right] \, .
\label{eq:dtRav}
\end{eqnarray}%
\end{subequations}%
The products $\Average{   R^{-1}  } {\Rav}$
and $\Average{   R^{-2}  } {\Rav}^2$ eventually approach
one because the size distribution becomes monodisperse in the long-time
limit. Hence, in this limit the characteristic aggregate volume,
$(4\pi/3)\,{\Rav}^3$, follows exactly the
same law, \Eq{avRcube}, as the growth of the average aggregate volume 
$(4\pi/3)\,\Average{ R^3 }$, 
\begin{eqnarray}
  {\Rav}^2 \: \diffT{{\Rav}}
  &=& \sigma D \, (k-1) 
  \quad \text{for large } t \, .
  \label{eq:Rbar-growth}
\end{eqnarray}
This is demonstrated in \Fig{RbarCubed} by showing that the ratio
${\Rav}^2 \diffT{{\Rav}} / [\sigma D \, (k-1)]$ settles to one after
some initial transient.  
In order to also understand the transient decay to the growth law,
\Eq{Rbar-growth}, we take a closer look at the difference of the time
evolution of $\Average{ R }^3$ and $\Average{ R^3 }$.

\subsection{Deviation of $\Average{ R }^3$ from $\Average{ R^3 }$}

Equations \eq{avRcube} and \eq{Rbar-growth} state that in the long run
the expectation values $\Rav^3$ and $\Average{ R^3 }$ acquire the
same slope as functions of time. In order to gain insight into the
difference of the two functions, we consider the expectation
value~\Average{ R^4 }.

We use $R = \Rav + (R-\Rav)$ and the forth power of this expression to observe that
\begin{widetext}
\begin{eqnarray}
  \Average{ R^4 } - \Average{ R^2 }^2
&=& - \left(  \Average{ R^2 } + \Rav^2 \right) \; \Average{ \left( R - \Rav \right)^2 }
+  6\: \Rav^2 \; \Average{ \left( R - \Rav \right)^2 }
+  4\: \Rav \;   \Average{ \left( R - \Rav \right)^3 }
+              \Average{ \left( R - \Rav \right)^4 }
\nonumber\\[2mm]
& = & 4\: \Rav^2 \; \Average{ \left( R - \Rav \right)^2 } \;
\left[
  1 + \frac{\Average{ \left( R - \Rav \right)^3 }}{\Rav \; \Average{ \left( R - \Rav \right)^2 }} 
    + \frac{\Average{ \left( R - \Rav \right)^4 }}{4 \; \Rav^2 \; \Average{ \left( R - \Rav \right)^2 }} 
    -      \frac{\Average{ \left( R - \Rav \right)^2 }}{4\; \Rav^2} 
\right]
\label{eq:R4average}
\end{eqnarray}%
When approaching a monodisperse distribution the expression in square
brackets rapidly approaches one, with corrections of order
$\Rav^{-2}$.  This observation provides the following insight into the
leading order contributions to the difference $\Average{R^3}-\Rav^3$,
\begin{eqnarray*}
  \Average{ R^3 } 
&=& \Average{ \left[ \Rav + \left( R - \Rav \right) \right]^3}
 \simeq  \Rav^3 + 3 \Rav \; \Average{ \left( R - \Rav \right)^2 }
 \simeq  \Rav^3 + \frac{3}{4 \: \Rav}  \; \Average{ \left( R^2 - \Average{ R^2 } \right)^2 }
\end{eqnarray*}%
\end{widetext}%
where we used \Eq{R4average} in the last step.  Rearranging the
equation yields
\begin{subequations}
\begin{eqnarray}
\Rav^3 &=& \Average{ R^3 } - \frac{ 3 \Omega_2 }{ 4 \: \Rav }
\label{eq:R3average}
\\[1mm]
\text{with \qquad} \Omega_2 &=& \Average{ \left( R^2 - \Average{ R^2 } \right)^2 }  \, .
\label{eq:defOmega}
\end{eqnarray}%
\end{subequations}%
\begin{figure}%
{\footnotesize
  \input{fig4_Rbar_cube_evolution.tex}
}
\caption[]{Evolution of ${\Rav}^2 \diffT{{\Rav}}$ for different values
  of $k$, as indicated in the legend.  
  The data is obtained by evaluating \Eq{dtRav} for our numerical data.
  As predicted by \Eq{Rbar-growth} it always approaches $\sigma D \, (k-1)$ for large $t$. 
  In the inset we show the mismatch of the numerical data and the improved prediction, 
  \Eq{RcubeCorrectionFull}.
  \label{fig:RbarCubed}}
\end{figure}%
Numerical data shows that $\Omega_2$ has a much weaker time dependence than $\Rav^{-1}$.
Hence, the time derivative of \Eq{R3average} amounts to
\begin{subequations}
\begin{eqnarray}
  \Rav^2 \; \frac{\rmd}{\rmd t} \Rav 
& \simeq &  \sigma D \, (k-1)
 + \frac{\Omega_2}{4 \: \Rav^4} \; \Rav^2 \; \frac{\rmd}{\rmd t} \Rav 
\nonumber \\[2mm]
\Leftrightarrow 
\Rav^2 \; \frac{\rmd}{\rmd t} \Rav 
&=& 
 \sigma D \, (k-1) \; \left( 1 - \frac{\Omega_2}{4 \, \Rav^4} \right)^{-1} \, .
\label{eq:RcubeCorrection}
\end{eqnarray}
The dotted grey line in \Fig{RbarCubed} shows the resulting prediction
when one assumes that $\Omega_2$ never noticeably deviates from its
initial value
\begin{equation}
\Omega_2 
= \frac{1}{5} \; \frac{\Rmax^5-\Rmin^5}{\Rmax-\Rmin} 
- \left( \frac{1}{3} \; \frac{\Rmax^3-\Rmin^3}{\Rmax-\Rmin} \right)^2
\end{equation}%
\label{eq:RcubeCorrectionFull}%
\end{subequations}%
determined for the initial aggregate size distribution, \Eq{IC}.
For the specified values $\Rmax=3$ and $\Rmin=0.02$ it takes the value
$\Omega_2 \simeq 7.19$.
The inset of \Fig{RbarCubed} shows the difference between this
prediction and the numerical data.  The strong fluctuation in the data
for $k\lesssim 5$ are due to singularities in the evolution arising
when an aggregate reaches zero radius. They reflect the evaporation of
aggregates, and we will not apply \Eq{RcubeCorrectionFull} in that
case since it was derived based on the assumption of no
evaporation. On the other hand, for $k \gtrsim 5$ and $t-t_0 \gtrsim
t_0$, \ie once the overall aggregate volume has doubled,
\Eq{RcubeCorrectionFull} provides an accurate description of the
evolution.

\subsection{The variance of the distribution}
\label{sec:2nd-cumulant}

Equations \eq{R4average} and \eq{defOmega} provide the variance of the
aggregate size distribution
\begin{equation}
  \Average{ \left(R - \Rav \right)^2}  \simeq \frac{\Omega_2}{4 \: \Rav^2} \, .
\label{eq:variance}
\end{equation}
Remarkably, the standard deviation decays like $\Rav^{-1}$.  Based on
the approximation that the aggregate size distribution amounts to a
Gaussian at all times this results has previously been obtained by
\citet{ClarkKumarOwenChan2011}. However, in contrast to \Eq{variance}
these authors predicted a slightly different decay that scales like
$\Rav^{-2+2/(k-1)}$.  In \Sec{stretching} we will show that this
discrepancy arises from a very slight time dependence of
$\Omega_2$: it increases like $\Rav^{2/(k-1)}$.  For large $k$
this correction is negligible such that it is not captured by the
present analysis.

The central results of this section are \Eqs{Rbar-growth} and
\eq{RcubeCorrection}. 
They express that one can accurately integrate the average radius
\Rav\ without need to refer to the evolution of the individual
aggregates: the average \Rav\ need not be calculated self-consistently
as an average over the aggregates, but it has its own evolution
equation, \Eq{RcubeCorrection}.  The solution of this equation
explicitly solves the global constraint that couples the set of
equations \eq{Ri_evolution}.  Hence, the $N$ dimensional system of
non-linear coupled equations \eq{Ri_evolution} for the aggregate radii
$R_i$ is reduced to $N$ identical one-dimensional differential
equations that only differ by their initial conditions.
Henceforth, we concentrate on this equation and suppress the index $i$.

\mysection{The reduced aggregate radius}
\label{sec:reduced-size}

The evolution of the decoupled set of equations \eq{Ri_evolution} is
most conveniently studied based on the reduced aggregate radius $\rho
= R / {\Rav}$ that accounts for the trivial drift of the aggregate
size due to the overall volume growth.

\subsection{Equation of motion}

\begin{figure}
{\footnotesize
  \input{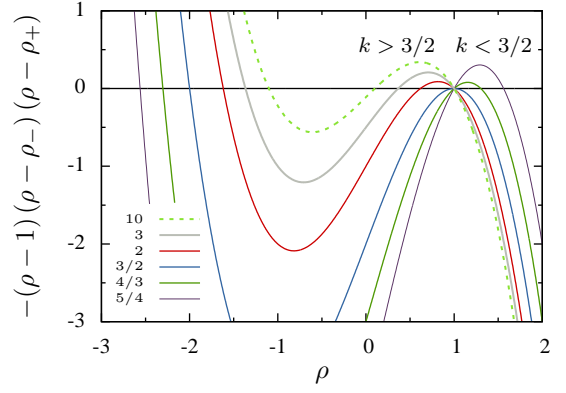}
}
\caption[]{The cubic polynomial in the numerator of
  \Eq{rhodotasympt}. For all $k>1$ its three roots give rise to three fixed points
  of the reduced radius, $\rho$ that are located at $\rho=1$ and $\rho=\rho_{\pm}$. 
  For $k=3/2$ there is a bifurcation where the
  roots $\rho=1$ and $\rho_+$ 
  change stability. 
  \label{fig:rhoDot}}
\end{figure}

Using \Eq{Ri_evolution} the time derivative of $\rho$ can be written as
\begin{eqnarray}
\dot \rho
  &=&   
  \frac{d}{dt} \frac{R}{{\Rav}} 
  = \frac{\dot R}{{\Rav}} - \rho \; \frac{\diffT{{\Rav}}}{{\Rav}}
  \nonumber
  \\[2mm]
  &=& 
  \frac{\sigma D}{{\Rav} \; R^2} \;  \left[ k \; \rho -1 \right]
  - \rho \; \frac{\diffT{{\Rav}}}{{\Rav}}
  \nonumber
  \\[2mm]
  &=& 
 - \frac{\sigma D}{{\Rav}^3} 
 \; \rho^{-2} 
 \; \left[ 
   \frac{{\Rav}^2 \, \diffT{{\Rav}}}{\sigma D} \; \rho^3 - k \; \rho + 1
   \right]
\label{eq:rhodot}
\end{eqnarray}
According to \Eq{Rbar-growth} (or \Fig{RbarCubed}) the factor 
${\Rav}^2 \, \diffT{{\Rav}} / (\sigma D)$ 
approaches $k-1$ after
a short initial transient. 
Consequently,
\begin{subequations} 
\begin{eqnarray}
\dot \rho 
  & \simeq &   
  - \frac{\sigma D}{{\Rav}^3} 
 \; \frac{
   (k-1) \; \rho^3 - k \; \rho + 1
   }{ \rho^{2} }
  \nonumber
  \\[2mm]
  &=& 
  -\frac{\sigma D \, (k-1)}{{\Rav}^3} \; 
  \frac{    \left( \rho - 1 \right) \;
    \left( \rho - \rho_-  \right) \;
    \left( \rho - \rho_+  \right) }{ \rho^2 }
  \label{eq:rhodotasympt}
\end{eqnarray}
with 
\begin{equation}
  \rho_\pm (k) = -\frac{1}{2} \pm \frac{1}{2}\; \sqrt{ \frac{k+3}{k-1} } \, .
  \label{eq:rho-pm}
\end{equation}%
\label{eq:rhodotasympt-all}
\end{subequations}%
The right-hand side of \Eq{rhodotasympt} involves a cubic polynomial
in $\rho$ (\Fig{rhoDot}). For all $k>1$ it gives rise to three fixed
points of the reduced radius: the average aggregate radius $\rho=1$, a
non-trivial radius $\rho_+$, and an unphysical fixed point $\rho_-$ at
negative values of $\rho$. Discussing their positions and stability
for different reduced temperature ramp rates, \Fig{phase_flow},
provides detailed insight into the dynamics.
\begin{description}
\item[$k=1$] \ We recover classical Ostwald ripening. The radius
  $\rho_+$ diverges, and the constraint on the overall aggregate
  volume gives rise to an asymptotic aggregate size distribution where
  the largest aggregates are of radius $\rho_{\textrm{max}}=3/2$.
\item[$1 < k < 3/2$] \ \Eq{rhodotasympt} has an unstable fixed
  point at $\rho=1$, \ie for $R = {\Rav}$. Aggregates that are smaller
  than the average radius shrink and they evaporate eventually when
  they reach the radius $\rho=0$.  Aggregates larger than $\rho_+$
  shrink, too, until they reach the stable aggregate radius $\rho_+$.
  On the other hand, aggregates in the range of $1<\rho<\rho_+$ grow
  at the expense of the shrinking ones, also striving to reach the
  aggregate radius $\rho_+$.
  When all aggregates are smaller than $\rho_+$ and $\rho_+ \gg 3/2$
  we expect a similar dynamic scaling theory to be applicable as the
  Lifshitz-Slyozov description of Ostwald ripening for $k=1$ 
  \cite[see ][ for some pioneering work discussing this situation]{Slezov2009}. 
  In the following we concentrate on the case $k>3/2$.
\item[$k=3/2$] \ The fixed points $\rho=1$ and $\rho_+$ cross, and
  they exchange their stability. Beyond this value aggregate evaporation
  ceases wen all remaining aggregates have a size $\rho>\rho_+$.
\item[$k > 3/2$] \ \Eq{rhodotasympt} has a stable fixed point for
  $\rho=1$, and an unstable fixed point at $\rho_+$ which rapidly
  approaches $k^{-1}$ for $k \gtrsim 5$. 
  After a brief initial transient no aggregates evaporate any
  longer, and the distribution becomes strongly peaked around the
  average aggregate radius ${\Rav}$. This is indeed what we
  have observed in \Fig{CDFevolution}. 
\end{description}

\subsection{Evaporation of aggregates}
\label{sec:evaporation}

\begin{figure}
{\footnotesize
  \input{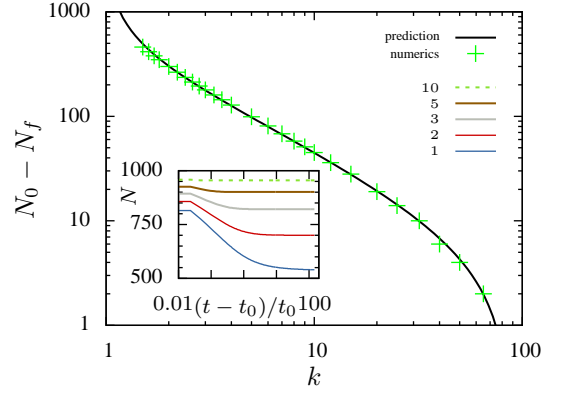}
}
\caption[]{Evolution of the aggregate number. The inset shows the
  time evolution of the number of aggregates for different values of $k$.
  All systems are initialised with $N_0 = 1000$ aggregates with a size
  distribution as outlined in \Eq{IC}. Eventually, they settle down to
  a constant aggregate number, $N_f$. 
  The main panel compares the number of evaporated aggregates 
  $N_0 - N_f$ to the prediction that it should amount to  
  $\int_0^{\rho_+} n(\varrho,t=t_0)\; \rmd\varrho$. 
  \label{fig:aggregate_number}}
\end{figure}

Aggregates that are smaller than ${\Rav}$ by a factor of $\rho_+$ shrink
and evaporate when they reach zero size. For large values of $k$ and
reasonably smooth initial aggregate densities this can only be a small
fraction of aggregates. Consequently, $n$ does not change much
when these aggregates disappear.
To support this view we show in \Fig{aggregate_number} that to an
excellent approximation the number of aggregates bound to evaporate
amounts to the number of aggregates in the initial distribution that lie
below $\rho_+$.

The fate of a general initial distribution for an initial value of $k$
in the range $1 < k \leq 3/2$ can be discussed based on
\Fig{phase_flow}. For $1 < k \leq 3/2$ the aggregates with a radius smaller
than average shrink, and eventually they evaporate. While doing so the
number density, $n$, decreases. According to \Eq{define-k}
this results in an increase of $k$. This growth of $k$ continues until all
aggregates have a size $\rho > \rho_+$, \ie their size lies above the
the red line in \Fig{phase_flow}.  At that time $k$ takes a value
$k \gtrsim 3/2$, and in the subsequent long-time limit, $k$ is a
constant of motion.

For the initial conditions specified by \Eq{IC} no aggregates should
evaporate for $\Rmin/\Rav > \rho_+(k_c) \simeq k_{c}^{-1}$, \ie
for $k_c > 75$.  In practice, the numerical simulations show that
$k_c$ is slightly smaller.  Systems subjected to a temperature
ramp where $k > 64$, \ie for $\xi \gtrsim 250 \pi\sigma D \, n$ evolve
at a constant number density, $n$, of aggregates, and hence at a
constant value of $k$.
When dealing with numerical data we always indicate the initial value
of $k$, and self-consistently take into account its change in the
plots.
Our focus of attention will be the asymptotics of the shape of the
aggregate size distribution.

\subsection{Evolution of the reduced aggregate radius}

For all $k \gtrsim 3/2$ and sufficiently late times the evolution of
the reduced aggregate radius, $\rho$, 
can be determined explicitly by integrating \Eq{rhodot}.
Introducing the function 
\begin{equation}
a = \Rav^3 / \left[ 3\sigma D \, (k-1) \right]
\label{eq:define-a}
\end{equation}
and focusing on values $\rho \simeq 1$ we write
\begin{subequations}
\begin{eqnarray}
  3 \, (k-1)\, a \; \rho^2 \; \dot \rho
  & = &   
  - (k-1) \: \dot a\, \rho^3 + (k-1) \rho + (\rho - 1)
\nonumber \\[2mm]
  & \simeq &
  - (k-1) \; \rho \; \left[ \dot a \, \rho^2 - 1 \right] 
\label{eq:rhodot-withA-evolution}
\\[2mm]
\Leftrightarrow \qquad
  \frac{2}{3} \; a^{-1/3}
  & = &
  \frac{\rmd}{\rmd t} \left( a^{2/3} \rho^2 \right) \, .
\label{eq:rhodot-withA}
\end{eqnarray}%
\label{eq:rhodot-withA-full}%
\end{subequations}%
This equation allows us to evaluate the derivative
\begin{subequations}
\begin{eqnarray}
\frac{\rmd}{\rmd t} R^2 
&=& \left[ 3 \, \sigma D \, (k-1)\right]^{2/3} \; \frac{\rmd}{\rmd t}\left( a^{2/3} \, \rho^2 \right) 
\nonumber\\[2mm]
&=& \frac{ 2 \,\sigma D\, (k-1)}{\Rav}
\label{eq:ddtR2}
\end{eqnarray}
which agrees with the time derivative of $\Average{R^2}$ up to a tiny correction
\begin{eqnarray}
\frac{\rmd}{\rmd t} \Average{ R^2} 
&=& \Average{ 2 R \; \dot R } 
\nonumber \\
&=& \frac{ 2 \,\sigma D\, (k-1)}{\Rav} \; \left[ 1 + \frac{1 - \Rav \, \Average{R^{-1}}}{k-1} \right] \, .
\label{eq:ddtR2av}
\end{eqnarray}%
\end{subequations}%
Altogether, \Eqs{ddtR2} and \eq{ddtR2av} imply that
\begin{equation}
\frac{\rmd}{\rmd t} \left( R^2 - \Average{ R^2 } \right) = 0 \, .
\label{eq:R2evolution}
\end{equation}
After all, there can be no merely time-dependent function appearing on the
right-hand side of this equation because the expectation value
\Average{R^2 - \Average{R^2}} must vanish at any time.

The result, \Eq{R2evolution}, states that at late times aggregates always grow in such a
way that the difference, $R^2 - {\Rav^2}$, is preserved.
This has immediate implications on the aggregate size distribution
which will be discussed in the next section.

\begin{figure*}
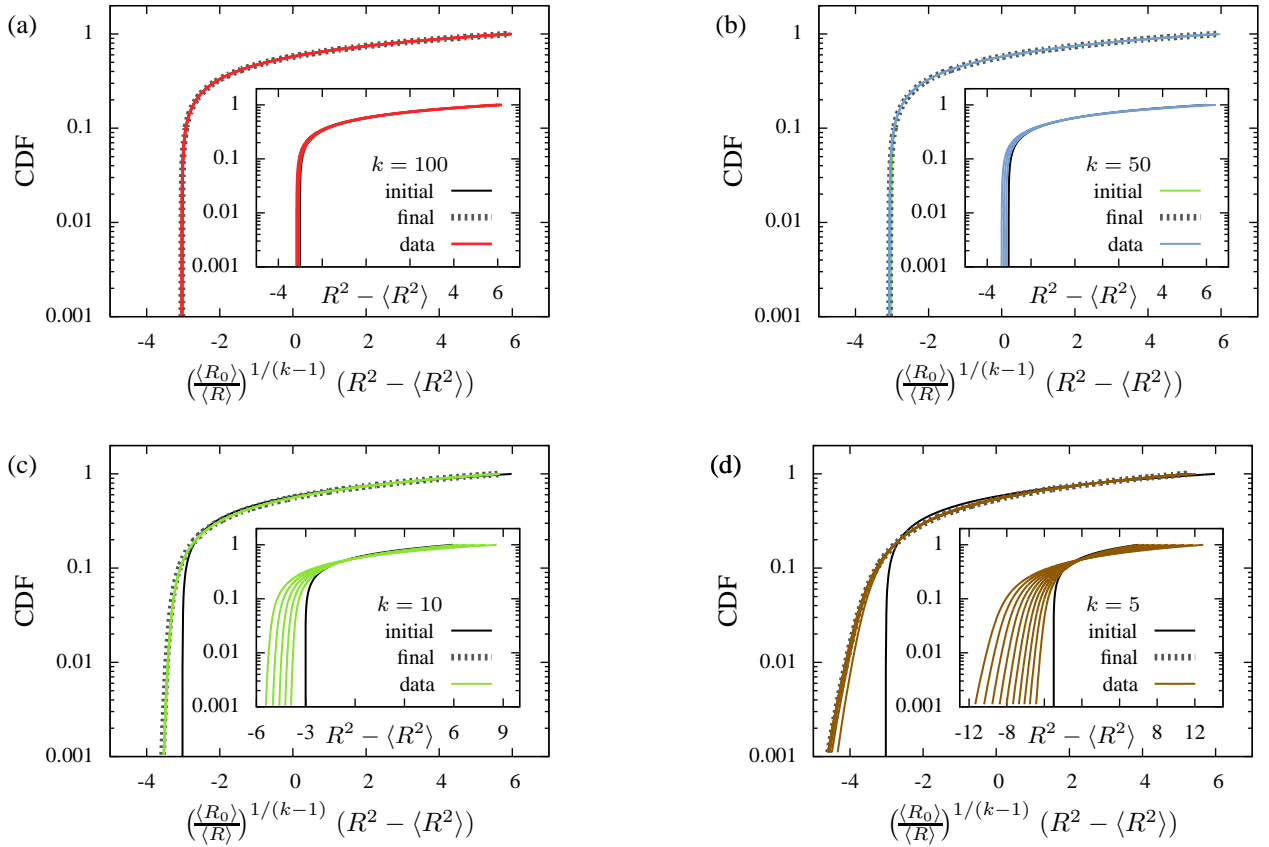

{\footnotesize
\begin{minipage}{0.48\textwidth}
  \input{fig7_cdfS__k_0100.tex}
\end{minipage}
\hfill
\begin{minipage}{0.48\textwidth}
  \input{fig7_cdfS__k_0050.tex}
\end{minipage}
}
\\[3mm]
{\footnotesize
\begin{minipage}{0.48\textwidth}
  \input{fig7_cdfS__k_0010.tex}
\end{minipage}
\hfill
\begin{minipage}{0.48\textwidth}
  \input{fig7_cdfS__k_0005.tex}
\end{minipage}
}
\caption[]{The CDFs (a) $k=100$, (b) $k=50$, (c) $k=10$, and (d)
    $k=5$.  The insets show the CDFs as a function of $x = R^2
  -\Average{R^2}$ as suggested in \Eq{defX}, and the main panels the
  dependence on $\tilde{x}$ which has been defined in
  \Eq{def-tilde-x}.  
  The initial conditions are provided by solid black labelled by
    the tag ``initial'', and thin lines with colours matching those in
    \Fig{CDFevolution} (``data'') show the numerical results for a
  progression of time on a logarithmic scale. We provide here data
    where \Rav\ grows to a size $\Rav \simeq 100$.  As function of $x$
    the CDFs are slowly broadening in time (insets). As function of
    $\tilde x$ they approach an asymptotic distribution (``final'') that is
    provided by dotted grey lines in the main panels.
  \label{fig:size_distribution}}
\end{figure*}

\mysection{Evolution of the aggregate size distribution}
\label{sec:distribution}

According to \Eq{sortedR} the order of the aggregate radii is
preserved by the dynamics: when aggregate $i$ is smaller than
aggregate $j$ initially, this holds also at all later times.
Based on this observation and the explicit integration of the
evolution equation of the aggregate radius, \Eq{R2evolution}, one can
derive the aggregate size distribution.
This is most easily done based on the cumulative aggregate size
distribution function (CDF).

\subsection{Initial distribution, and its evolution based on \Eq{R2evolution}}

For convenience of the discussion of the asymptotic shape of
the CDF, we immediately remove the aggregates from the initial
distribution that will evaporate. 
According to the arguments underpinned by \Fig{aggregate_number} this
amounts to the aggregates smaller than $R_c = \Average{ R_0 } \: \rho_+(k)$,
where  $\Average{ R_0 } = (\Rmax + \Rmin)/2 = 1.51$ is
the average radius with respect to the initial aggregate size
distribution \eq{IC}.
When no aggregates evaporate we set $R_c = \Rmin$. 
With this adoption, the CDF characterising the initial distribution,
$\CDF (R_0)$, takes the form
\begin{equation}
\CDF (R_{0}) = \left \{ 
  \begin{array}{lll}
  0  
  & \text{ for }
  & \qquad R_0 < R_c \, , 
  \\[2mm]
  \frac{ R_0 - R_c }{ \Rmax - R_c }
  & \text{ for }
  & R_c < R_0 < \Rmax \, ,
  \\[2mm]
  1  
  & \text{ for }
  & \Rmax < R_0 \, .
\end{array} 
\right. 
\label{eq:CDF0}
\end{equation}
To avoid the involved notation required to explicitly distinguish the
different branches of the function, we henceforth only specify
its non-trivial branch, and keep in mind that the function should be
set to zero when the expression drops below zero, and set to one when it
rises beyond one. 

In order to apply \Eq{R2evolution} it is
convenient to rewrite \Eq{CDF0} as a function of 
\begin{equation}
x = R^2 - \Average{R^2} \stackrel{!}{=} R_0^2 - \Average{ R_0^2 } \, .
\label{eq:defX}
\end{equation}
In that case the non-trivial branch of the CDF takes the form of a
square-root dependence
\begin{eqnarray}
   \CDF ( x ) 
   &=& \frac{ \left[ x + \Average{ R_0^2 } 
              \right]^{1/2} - R_c }
            { \Rmax - R_c }
   \label{eq:CDF}
\end{eqnarray}
The initial condition $\mathcal{C}(x)$ of the CDF, provided as a
function of $x$, is shown by solid black lines in
\Fig{size_distribution}(inset).
To determine the time dependence of the CDF we note that according to
\Eq{defX} the value of $x$ is preserved during the evolution.
Consequently, the CDF should not change in time when it is plotted as
a function of $x$.
To test this assertion the insets of \Fig{size_distribution} show the
initial conditions together with the CDF at later times, that are
shown in colours matching those used in \Fig{CDFevolution}. Except for
the change of the variable, $x$ rather than $R$, the CDFs shown in
\Fig{CDFevolution} and \Fig{size_distribution}(inset) differ only by a
different choice of the time increments.  A larger factor of overall
volume growth has been chosen in \Fig{size_distribution} in order to
display distributions where the average radius grows to $\Rav \simeq
100$ rather than only till $9$.

The prediction that the CDF remains invariant, \Eq{CDF}, when plotted
as a function of $x$ properly captures main features of the time
evolution: the CDFs fall on top of each other and they tend to
preserve their form when plotted as a function of 
$ x = R^2 - \Average{ R^2 } $.
For all $k \gtrsim 50$ this provides an accurate description of
the numerical data.  On the other hand, for decreasing $k$ the tails
of the distributions towards the smaller aggregate sizes tend to
become less steep, and in addition there is a noticeable broadening of
the distributions in the course of time.
These deviations arise from the fact that for $\rho\simeq 1$ we
systematically underestimates the slope of $\dot\rho$ due to
suppressing the term $(\rho-1)/(k-1)$ on the right hand side of
\Eq{rhodot-withA-evolution}.

\subsection{Accounting for broadening and shape changes}
\label{sec:stretching}

For late times, where \Eq{Rbar-growth} applies, we can gain insight
into the broadening of the distribution by integrating
\Eq{rhodotasympt-all} rather than \Eq{rhodot-withA-full}.

We use \Eq{define-a} to write \Eq{rhodotasympt} in the form
\begin{equation}
\dot \rho = - \frac{1}{3 \, a}
  \frac{    \left( \rho - 1 \right) \;
    \left( \rho - \rho_-  \right) \;
    \left( \rho - \rho_+  \right) }{  \rho^2 }
\label{eq:dotRhoWithA}
\end{equation}
and introduce a function $g(\rho)$ that obeys the differential
equation 
\begin{equation}
\frac{\rmd g}{\rmd \rho}
= \frac{ \rho^2 \; g }
  { \left( \rho - 1 \right) \;
    \left( \rho - \rho_-  \right) \;
    \left( \rho - \rho_+  \right) } 
\label{eq:dgdrho}
\end{equation}
Combining \Eqs{dotRhoWithA} and \eq{dgdrho} 
allows us to rephrase the evolution of $\rho$ in the form
\begin{eqnarray}
\frac{\dot g}{g} &=& g^{-1} \; \frac{\rmd g}{\rmd \rho} \; \dot \rho 
= \frac{-1}{3 \: a}  = - \frac{\dot a}{3 \, a} \, ,
\label{eq:ode-g-a}
\end{eqnarray}
where we used in the last step that $\dot a = 1$ in the long-time
asymptotics considered here.
Equation \eq{ode-g-a} implies that 
\begin{equation}
\frac{\rmd}{\rmd t} \left( g \; a^{1/3} \right) = 0 \, .
\label{eq:stretchSolution}
\end{equation}
In order to interpret this finding we have to find the function~$g$.
The differential equation \eq{dgdrho} has solutions of the form
\begin{subequations}%
\begin{eqnarray}%
  g &=& C\;
      (\rho-1)^{\alpha_1} \; 
      (\rho -\rho_-)^{\alpha_-} \;
      (\rho-\rho_+)^{\alpha_+} \,,
\label{eq:xArg}      
\label{eq:stretching-ansatz}%
\end{eqnarray}%
where the constant number $C$ represents the integration constant.
Inserting \Eq{xArg} into \Eq{dgdrho} provides a
linear set of equations for the exponents $(\alpha_1, \alpha_-, \alpha_+)$
that is solved by
\begin{eqnarray}%
  \alpha_1 &=& \frac{1}{(2+\rho_+) \, (2+\rho_-)} 
  = \frac{k-1}{2\,k-3} \, ,
  \\[2mm]
  \alpha_- &=& \frac{ \rho_-^2}{(2+\rho_+) \; (1 + 2\rho_+)}
  \simeq  \frac{1}{2} - \frac{1}{4k} + \frac{5}{8k^2} - \dots \, ,
  \\[2mm]
  \alpha_+ &=& \frac{ \rho_+^2}{ (2 + \rho_- ) \; (1 + 2\rho_-) }
  \simeq  - \frac{1}{k^2} + \dots \, ,
\end{eqnarray}
\end{subequations}%
Equation \eq{stretchSolution} together with the definition of $a$, \Eq{define-a}, entails that
the cumulative distribution function is a function of
$\Rav\,g$. 
Moreover, the insets of \Fig{size_distribution} show
that in leading order of the long-time asymptotics, where
$\Average{R^2} = \Rav^2$ (\cf\Eq{variance}), the cumulative
distribution function must depend on $R^2 - {\Rav}^2 = {\Rav}^2 \:
(\rho^2 - 1)$.
This dependence can be faithfully recovered from $(\Rav\,g)^{1/\alpha_1}$ 
by observing that 
$\alpha_1^{-1} = 2-(k-1)^{-1}$.
Moreover, making use of $\alpha_1 + \alpha_+ + \alpha_- = 1$ one easily shows that  
$\alpha_-/\alpha_1 = 1 -(k-1)^{-1} - \alpha_+/\alpha_1$. 
These relations provide 
\begin{widetext}
\begin{subequations}
\begin{eqnarray}
\left( \Rav \, g \right)^{\alpha_1^{-1}}  
&=& \Rav^{2-(k-1)^{-1}} \;
  ( \rho^2 - 1 ) \; 
  \left( 1 + \frac{\rho_+}{\rho+1} \right) \; 
  \left( \rho + 1 + \rho_+ \right)^{-(k-1)^{-1}} \;
  \left( \frac{ \rho-\rho_+ }{  \rho + 1 + \rho_+ } \right)^{\alpha_+/\alpha_1}
\label{eq:stretched-x-full}%
\\[2mm]
&\simeq&
   \Rav^{-(k-1)^{-1}} \; \left( R^2 - \Rav^2 \right) 
   \; \left[ 1 + \mathcal{O}\!\left( (k-1)^{-1} \right) \right] \, .
\label{eq:stretched-x}%
\end{eqnarray}%
\end{subequations}%
\end{widetext}%
The factor $ \Rav^{-1/(k-1)}$ in \Eq{stretched-x} entails that
$\mathcal{C}( R^2 - \Average{R^2} )$ features a sustained broadening,
as observed for the CDFs shown in the insets of
\Fig{size_distribution}. In line with the $k$ dependence of this
factor the broadening is increasingly more pronounced for smaller
values of $k$.
In contrast the CDFs should remain invariant when accounting of the
broadening by plotting $\mathcal{C}$ as a function of
\begin{equation}
  \tilde x 
= \left(
     \frac{  \Average{R_0}  }{ \Rav } 
  \right)^{(k-1)^{-1}} \; 
  \left( R^2 - \Average{R^2}  \right)
\label{eq:def-tilde-x}
\end{equation}
This variable accounts for the sustained broadening
of the CDF via the factor $\Rav^{-1/(k-1)}$, and at early times it
appropriately fixes the mean position of the CDF, as observed in
\Eq{defX}.

The data collapse of the CDFs shown in the main panels of
\Fig{size_distribution}(a) and (b) demonstrates that for $k \gtrsim
50$ the CDFs are invariant when plotted as a function of $\tilde x$.
For smaller values of $k$ the variable $\tilde x$ faithfully accounts
for the broadening of the distribution that was severely
underestimated previously. However, the higher-order corrections
specified by the last three factors in \Eq{stretched-x-full} affect
the relation between $R^2 - \Average {R^2}$ and its initial value
$R_0^2 - \Average {R_0^2}$ such that the shape of the distribution is
no longer be preserved (\Fig{size_distribution}(c) and (d)).
The dotted grey lines show the shape of the distribution that results
when these factors are accounted for.  Taking into account these terms
provides a parameter free prediction of the asymptotic form of the CDF
that is accurate for all considered values of $k$.

\subsection{Scaling of the centred moments of the size distribution}

The observation that the aggregate size distribution is invariant when
plotted as a function of $\tilde x$ has immediate consequences for the
centred moments of the size distribution function.
The data collapse implies that $\Average{ \tilde x^n}$ is invariant in time such that
\begin{equation}
\Omega_n := \Average{\left( R^2 - \Average{R^2} \right)^n }
 \sim  \left( \frac{ \Rav }{  \Average{R_0}  }
  \right)^{n/(k-1)} \, .
\label{eq:defOmega-n}
\end{equation}
For small $k$ the factor $\Rav^{2/(k-1)}$ provides a
small, but noticeable growth of $\Omega_2$ that is reflected in the
broadening of the distributions shown in the insets of \Fig{size_distribution}. 

In order to calculate the centred moments we note that
\begin{eqnarray*}
R - \Rav 
&=& \frac{ \left( R^2 - \Average{R^2} \right) 
  - \left( R - \Rav \right)^2
  + \Average{ \left( R - \Rav \right)^2 }}{2\Rav} 
\\[2mm]
&=&  \frac{1}{2\Rav} \; 
\left[ \left( R^2 - \Average{R^2} \right) 
  + \mathcal{O}\left( \Rav^{-2} \right)
\right]
\end{eqnarray*}
Consequently, 
\begin{eqnarray*}
\Average{ \left( R - \Rav \right)^n } 
&\simeq& \Average{ \left( \frac{ R^2 - \Average{ R^2 } }{ 2 \, \Rav } \right)^n } 
=  \frac{ \Omega_n }{ \left( 2 \, \Rav \right)^{n} }
\end{eqnarray*}
In view of the asymptotic scaling, \Eq{defOmega-n}, of $\Omega_n$ this implies
\begin{equation}
  \Average{ \left( R - \Rav \right)^n } \sim  \Rav^{-n+n/(k-1)} \; .
 \end{equation}
In particular, we hence obtain the result anticipated in \Sec{2nd-cumulant}:
the standard deviation of the aggregate size distribution decays like
\begin{equation}
  \sqrt{ \Average{R^2} - \Rav^2 } 
=  \frac{{\Omega_2}^{1/2}}{2 \, \Rav} 
\sim \Rav^{-1+(k-1)^{-1}} \, .
\label{eq:std-deviation}
\end{equation}

\mysection{Discussion}
\label{sec:discussion}

The data collapse achieved in \Fig{size_distribution} and the resulting
scaling, \Eq{std-deviation}, of the standard deviation of the size
distribution underpin the assertion of \Sec{average-size} that the
aggregate size distribution tends to become monodisperse when
aggregates grow in an environment that leads to a sustained growth in
their net volume. For all $k \gtrsim 5$ we have provided a scaling form of the  
asymptotic shape of the size distribution, and for $k\gtrsim 50$ 
the initial condition is described so faithfully by this scaling form that
we have obtained a parameter-free prediction for all times.
In order to digest the relevance of these findings it is important to
estimate the order of magnitude of $k$ for different processes.

\begin{figure}
{\footnotesize
  \input{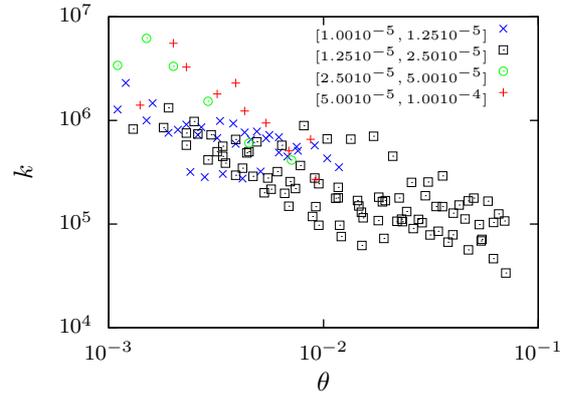}
}
\caption[]{The $k$-values for the water-rich phase of the
  water/iso-butoxyethanol mixture as a function of the reduced temperature
  $\theta = (T-T_c)/T_c$ where $T_c = 25.5^{\circ}$C is the critical temperature of the mixture. 
  Different symbols refer to measurements where the volume-fraction growth rates, $\xi$, take 
  values in the intervals indicated in the legend of the plot. 
  The values of $k$ have be calculated based on number densities $n$ reported by
  \citet{LappRohloffVollmerHof2012}, and the other material parameter
  are extrapolations of the respective parameters in literature which have been collected 
  by \citet{LappPhD2011}.
  \label{fig:k-values}}
\end{figure}

\subsection{Optical and calorimetric measurements on the phase separation of binary mixtures} 
\label{sec:phase-separation}

In a recent study \citet{LappRohloffVollmerHof2012} determined the
evolution of the number density, $n$, of droplets in the phase
separation of water/iso-butoxyethanol mixtures subjected to
temperature protocols that lead to a constant growth of the droplet
volume fraction.  The ramp rates $\xi$ of the increase of droplet
volume fraction ranged from $\xi = 10^{-5}$ to
$10^{-4}\;\text{s}^{-1}$.
Based on the temperature dependence of the pertinent material
parameters \cite{LappPhD2011} we show in \Fig{k-values} that in those
studies $k$ took values in the range of $10^{4} \dots 10^7$.  The wide
range of $k$ values results from exploring a range of ramp rates $\xi$
that covers one order of magnitude, and from the temperature dependence
of the material parameters.

Studies on other binary mixtures have observed phase separation during
a slow temperature ramp in differential scanning micro-calorimetry
\cite{HeimburgMirzaevKaatze2000,auernhammer05JCP} or by optical
measurements \cite{RullmannAlig2004,auernhammer05JCP}.  These
experiments typically involve gradual changes of the temperature $T$ by about
$1\;$K/h, which amounts to $\xi$ in the range also explored by
\cite{LappRohloffVollmerHof2012}.  Hence, we expect that they involve
similarly large values of $k$.

\begin{table}
\centering
\begin{tabular}{l@{$\qquad$}c@{$\qquad$}c@{$\qquad$}c}
\hline\\[-3mm]
  $T$ [$^\circ$C]   & $-40$ & $0$ & $10$ 
\\[1mm]
\hline\\[-2mm]
$\rmd \Phi/\rmd T$ \ [K$^{-1}$] & $2 \cdot 10^{-5}$ & $4.5 \cdot 10^{-4}$ & $8 \cdot 10^{-4}$
\\[1mm]
$\xi$ \ [s$^{-1}$]              & $2 \cdot 10^{-7}$ & $4.5 \cdot 10^{-6}$ & $8 \cdot 10^{-6}$
\\[1mm]
\hline\\[-2mm]
$\gamma$ \ [N/m]               & $8.4 \cdot 10^{-2}$ & $7.6 \cdot 10^{-2}$ & $7.4 \cdot 10^{-2}$
\\[1mm]
$\Phi$                       & $1.9 \cdot 10^{-4}$ & $6.1 \cdot 10^{-3}$ & $1.2 \cdot 10^{-2}$
\\[1mm]
$k$                            & $8.0$              & $5.7$              & $5.2$
\\[1mm]
\hline
\end{tabular}
\caption[]{Material constants for mixtures of water and air, and the
  resulting values for $\xi$ and $k$ based on a vertical wind speed of
  $\rmd H/\rmd t = 1\;$m/s. Values for other wind speeds can easily be
  obtained by observing that $k-1$ is proportional to $\rmd H/\rmd t$. 
  \label{tab:rain}}
\end{table}

\subsection{Growth of cloud droplets} 

Rain emerges when the air masses in a cloud rise due to topographic
constraints, or by encountering a cold front \cite{Mason1971,Rogers1989}. The
drop of pressure in response to the rising of height $H$ leads to
adiabatic cooling of the air. This in turn changes the solubility of water
in the air. Similarly to the phase separation discussed in
\Sec{phase-separation} this induces a continuous growth of cloud
droplets until they reach a size where collisions due to gravity and
inertia speed up their growth and trigger rain formation
\cite{BodenschatzMalinowskiShawStratmann2010}. \citet{Clement2008}
discussed the micro-physics of the droplet growth, emphasising the
importance of the heat of condensation and the impact of solutes in the droplets.

Here we augment his study by an estimate of the possible impact of
the continuous growth of the droplet volume fraction. We note that 
$\xi$ amounts to the product of three factors, 
\[
\xi = \frac{\rmd \Phi}{\rmd t} 
    = \frac{\rmd \Phi}{\rmd T} \; 
      \frac{\rmd T}{\rmd H} \; 
      \frac{\rmd H}{\rmd t} \, ,
\]
where $\Phi = V/\mathcal{V}$ is the volume fraction of droplets. 
The three factors on the right hand side of the equation 
amount to the slope of the phase boundary,%
\footnote{It has been demonstrated by \citet{LappRohloffVollmerHof2012}
  that ${\rmd \Phi}/{\rmd T}$ amounts to the slope of the binodal line
  of the phase diagram.}
${\rmd \Phi}/{\rmd T} \lesssim 5\cdot 10^{-4}\;\text{K}^{-1}$ 
\cite[p.~132]{MoranMorgan1997},
the  adiabatic lapse rate 
$\rmd T/\rmd H \simeq 1\,\mathrm{K}/100\,$m 
(\citep[p.~148]{MoranMorgan1997} or 
\citep[p.~29]{Rogers1989}),
and the average upwind speeds,
${\rmd H}/{\rmd t} = 1 \dots 10\;$m/s, respectively. 
This gives rise to values of $\xi$ between $5\cdot 10^{-6}$s$^{-1}$
and $5\cdot 10^{-5}$s$^{-1}$.

The number density of droplets in a cloud has been determined by
\citet{DitasShawSiebertSimmelWehnerWiedensohler2012} in recent
measurement campaigns, $n = 4.7\cdot 10^{8}$m$^{-3}$, and the values
of the diffusion constant and the Kelvin length are well-known
material constants. 
The latter is obtained by inserting the values of the
interfacial tension of the water-air interface, $\gamma$, the molar
volume of liquid water, $V_m = 18 \cdot 10^{-6}
\text{m}^3/\text{mol}$, \cite[p.~614]{Mason1971}, the equilibrium
volume faction of water vapour in air, $\Phi$, the molar gas constant,
$R = 8.3 \; \text{J/mol\,K}$, and the temperature $T$ into the
definition of the Kelvin length \cite{landau10}
\begin{equation}
\sigma =  \frac{2\, \gamma V_m \Phi}{R T} \, .
\label{eq:define-sigma}
\end{equation}
In \Tab{rain} we provide some representative data and the resulting
values for $\xi$ and $k$. 
For average vertical wind speeds of $1\;$m/s the values of $k$ lie in
the range of $5 \dots 8$, and for larger wind speeds higher values
are obtained.

We stress that the values provided in \Tab{rain} provide only a
rough, first order estimate of the parameters governing the evolution
of the droplet size distribution in clouds. 
Nevertheless, this estimate suggests that the droplet volume growth
due to the average rising of a cloud can give rise to values of $k$ in
the range, $k \gtrsim 5$ where the present results promise the arising
of interesting new physics.
This calls for a careful revisiting of the pertinent droplet growth laws.

\subsection{Synthesis of monodisperse colloids and nano-particles}

Fundamental work on the synthesis of monodisperse colloids goes back
to \citet{LaMerDinegar1950} and \citet{Reiss1951}. The theoretical
understanding of the mechanisms that lead to highly monodisperse
colloids and nano-crystals is still a topic of active research
\cite{RempelBawendiJensen2009,ClarkKumarOwenChan2011,SinghPuriDasgupta2012}.

For the synthesis of monodisperse silver particles (used for
photographic films) the material flux is well defined, and all
material constants required to determine the $k$-values have been
documented.
For the synthesis of Ag\,Br and Ag\,Cl particles \citet{Sugimoto1992} 
and \citet{SugimotoShibaSekiguchiItoh2000} provide
material constants and aggregate numbers that allow us to calculate
$k$ based on the increase of the total volume of the aggregates, $\xi
\mathcal{V}$, the diffusion coefficient $D$, and the Kelvin length~$\sigma$,
\begin{subequations}
\begin{eqnarray}
k &=& 1 + \frac{\xi}{4\pi D \sigma n} = 1 + \frac{Q_0 V_m}{4\pi D \sigma N} \, ,
\end{eqnarray}%
where $N$ is the number of aggregates in the sample volume $\mathcal{V}$, 
and 
\begin{eqnarray}
\xi  &=&  V_m \, Q_0 / \mathcal{V}
\end{eqnarray}%
is provided in terms of the molar volume, $V_m$, 
and the mass supply rate, $Q_0$.
Finally, the specific surface energy $\gamma$, 
the buffer temperature $T$, the mean-field monomer concentration $C_{\infty}$, 
and the molar gas constant $R = 8.314\;$J/(mol K) provide the Kelvin length as
\begin{eqnarray}
 \sigma &=& \frac{2\gamma V_m^2 C_{\infty}}{RT} \, .
\end{eqnarray}%
\label{eq:AgBr-k}
\end{subequations}%
\Tab{AgBr} provides the resulting $k$-values for 
different representative sets of $(T,D,C_{\infty},N)$ used for the synthesis of Ag\,Br particles,
and \Tab{AgCl} provides the $k$ values for the synthesis of
Ag\,Cl particles. Also in the latter case the $k$ values are obtained from
\Eqs{AgBr-k}, except that \citet{SugimotoShibaSekiguchiItoh2000}
provided the molar injection rate $q_0 = Q_0/\mathcal{V}$ and the number
density of particles, $n = N/\mathcal{V}$. 

\begin{table}
\centering
 \begin{tabular}{l@{$\quad$}c@{$\quad$}c@{$\quad$}c@{$\quad$}c}\hline\\[-3mm]
    $T$ [$^{\circ}$C] &   $40$            &   $50$             &    $60$            &   $70$    \\[1mm]
    $D$ [m$^2$/s] &  $9.94\cdot 10^{-10}$ & $1.26\cdot 10^{-9}$ & $1.56\cdot 10^{-9}$ & $1.92\cdot 10^{-9}$  \\[1mm]
$C_{\infty}$ [$\frac{\text{mol}}{\text{m}^3}$] & $1.01\cdot10^{-4}$& $2.12\cdot10^{-4}$ & $4.34\cdot10^{-4}$  & $8.42\cdot10^{-4}$   \\[1mm]
    $N$            & $3.20\cdot10^{17}$    & $1.25\cdot10^{17}$ & $4.60\cdot10^{16}$  & $2.20\cdot10^{16}$   \\[1mm]\hline\\[-2mm]
    $k$	           &    1.63              &   1.62              &  1.69              & 1.62 \\\hline
  \end{tabular}
\caption{Representative material parameters for the synthesis of monodisperse Ag\,Br particles 
\citep[adapted from][]{Sugimoto1992}
and the corresponding $k$ values as calculated via \Eqs{AgBr-k}. 
The molar volume of Ag\,Br is $V_m = 2.9\cdot 10^{-5}$m$^3$/mol, and its specific surface energy is $\gamma = 1.77\cdot10^{-1}$J/m$^2$. 
All experiments were conducted with a mass supply rate, $Q_0 = 10^{-3}$mol/s.
\label{tab:AgBr}}
\end{table}

The data show that the $k$ values selected for the synthesis of monodisperse silver
particles lie at $k\simeq 1.6$ for Ag\,Br-particles and in a range
between $6$ and $43$ for Ag\,Cl. 
Moreover, for the initial stages of the synthesis of Ce\,Sd nano-crystals
\citet{ClarkKumarOwenChan2011} estimated $k$ to lie in the range of 
$k \simeq 3 \dots 5$ (see their Fig.~4).
These choices have been obtained by tuning the temperature and the rates $Q_0$
or $q_0$ for optimal monodispersity of the product.  In all cases this
resulted in $k$ values larger than $3/2$ such that one can profit from
the size focusing arising for $k>3/2$. 
In principle, the values of $k$ should be chosen as large as possible
to achieve the smallest standard deviation, \Eq{std-deviation}, and
minimise the time required for the synthesis, \Eq{avRcube}.  In
practice, it becomes harder to realise stable and
reproducible experimental conditions for large values of $k$, and the
heat released in the growth might severely alter the present theory
for large growth rates.
Follow-up work will have to explore these effects.

\mysection{Conclusion}
\label{sec:conclusion}

In \Eqs{define-k} we have identified the dimensionless factor $k$ as
control parameter determining the features of the evolution of an
aggregate distribution evolving with overall volume growth. For $k=1$
(\ie no growth) the dynamics recovers the Lifshitz-Slyozov-Wagner
scenario of Ostwald ripening \cite{Voorhees1985,Bray1994}. For
$1<k<3/2$ we expect Ostwald-like behaviour as described by
\citet[Chap.~7]{Slezov2009}.
In the present paper we focused on the case $k>3/2$. On the one hand,
we established a new numerical algorithm, that is outlined in
\Fig{program_outline}. It allows us to accurately follow the evolution
of the aggregate size distribution over very long times because it
admits equidistant time stepping on a logarithmic time axis.
On the other hand, we have provided a complete analytical solution for
the evolution of the aggregate size distribution. It has no adjustable
parameters and agrees perfectly with the numerical data.

This excellent agreement establishes that for $k > 3/2$ the CDF
does not approach a scaling form.  Rather it is most conveniently
written as a function of the difference, $R^2 - \Average{R^2}$, of the
square of the considered radius, $R$, and its average, \Average{R^2}.  
We demonstrated in \Fig{size_distribution} that to a
very good approximation the shape of the distribution function remains
invariant when this dependence is augmented by a gradual broadening by
a factor $\Rav^{1/(k-1)}$.
Sub-dominant contributions to the evolution can arise from small
aggregates that grow slightly slower than those of average size. They
lead to noticeable changes in the small-size tail of the distribution
for $k \lesssim 10$. The resulting change of the shape of the
distribution can be accounted for by considering the higher order
correction in \Eq{stretched-x-full} and by self-consistently tracking
the influence of the evaporation of aggregates. The resulting
parameter-free prediction provides an excellent description of the
asymptotic shape of the distribution (dotted grey lines in
\Fig{CDFevolution}).  Consequently, the shape of the aggregate size
distribution is fully determined by its initial condition, rather than
by features of the dynamics.

\begin{table}
 \begin{tabular}{l@{$\quad$}c@{$\quad$}c@{$\quad$}c@{$\quad$}c}\hline\\[-3mm]
  $T$ [$^{\circ}$C] & 
  $25$	                 &  $30$             & $35$               &     $40$	\\[1mm]
  $D$ [m$^2$/s]   & 
  $1.44\cdot 10^{-9}$ & $1.64\cdot 10^{-9}$    & $1.86\cdot 10^{-9}$ & $2.11\cdot 10^{-9}$ \\[1mm]
  $C_{\infty}$ [$\frac{\text{mol}}{\text{m}^3}$] & 
  $5.04\cdot10^{-4}$      & $7.30\cdot10^{-4}$ & $1.04\cdot10^{-3}$  & $1.46\cdot10^{-3}$ \\[1mm]
  $n$[m$^{-3}$]  &
  $5.88\cdot10^{13}$      & $5.71\cdot10^{13}$ & $4.24\cdot10^{13}$  & $2.70\cdot10^{13}$ \\[1mm]
  $q_0$ [$\frac{\text{mol}}{\text{m}^{3}\text{s}}$] & 
  $5.95\cdot10^{-6}$      & $1.54\cdot10^{-5}$ & $3.86\cdot10^{-5}$  & $8.88\cdot10^{-5}$ \\[1mm]\hline\\[-3mm]
  $\xi$  [s$^{-1}$] &
  $1.54\cdot10^{-10}$     & $3.99\cdot10^{-10}$ & $1.00\cdot10^{-9}$  & $2.30\cdot10^{-9}$ \\[1mm] 
  $k$	&
  $6.26$                 &  $9.64$           & $19.4$             & $43.3$ \\[1mm]\hline
  \end{tabular}
  \caption{Material parameters for the synthesis of monodisperse Ag\,Cl particles \citep[adapted from ][table 3]{SugimotoShibaSekiguchiItoh2000}, and the resulting $k$-values as calculated via \Eqs{AgBr-k}. 
    For Ag\,Cl particles the molar volume is $V_m = 2.59\cdot 10^{-5}$m$^3$/mol, and their specific surface energy is $\gamma =
1.009\cdot10^{-1}$J/m$^2$. 
  \label{tab:AgCl}}
\label{AgCl}
\end{table}

In conclusion we have established that a weak thermal drift, or any other
mechanism that leads to slow aggregate growth, can have dramatic effects on
the aggregate size distribution. Even for very small effective driving
it has a noticeable impact on various features of the aggregate size
distribution.
\begin{itemize}
\item[The]\textbf{aggregate number density}
  is constant at late times (see \Fig{aggregate_number}).  In contrast
  to this finding for $k>1$, the ripening in isothermal systems (\ie
  for $k=1$) can only evolve by evaporation of small aggregates. This
  leads to a $t^{-1}$ decay of the number of aggregates.
\item[The]\textbf{mean aggregate radius} grows like $\Rav \sim t^{1/3}$. In contrast to
  Ostwald ripening, this growth is not connected to the evaporation of aggregates,
  but reflects the growth due to a constant volume flux onto the
  aggregates at a fixed number of aggregates. 
\item[The]\textbf{standard deviation of the aggregate radius}  decays with the non-trivial power
  $(t^{1/3})^{-1+1/(k-1)}$ (\cf\Eq{std-deviation}). Consequently, the
  relative width of the distribution, which amounts to the ratio of
  the standard deviation and the average radius, $\Rav$, 
  decays like $(t^{1/3})^{-2+1/(k-1)}$. The aggregate size
  distribution tends to become more and more monodisperse.
\item[The]\textbf{shape of the distribution} is governed by initial
  conditions, rather than being universal. When plotted as a function
  of $\tilde x$ specified by \Eq{def-tilde-x} the cumulative
  distribution function remains invariant except for small $k$ where
  there is a slight change of the tails.  It has been accounted for in
  the theoretical prediction shown by the dotted grey lines in
  \Fig{size_distribution}.
\end{itemize}
The latter two findings are in striking contrast to those of the
Lifshitz-Slyozov-Wagner theory of Ostwald ripening, which predicts that
the distribution approaches a universal distribution with a fixed
relative width.

For a range of different applications we have demonstrated in
\Sec{discussion} that values of $k > 3/2$, where these differences
prominently apply, may be regarded as common rather than as an
exception.  Consequently, the theory for the aggregate size
distributions, that we have established in \Sec{distribution}, opens
new opportunities in the characterisation and synthesis of aggregate
growth.
On the one hand, one can use the growth as a microscope to infer the
initial size distribution at nucleation from a measurement at a later
time when the aggregates have grown to a larger size. On the other
hand, the distinct dependence of the size distribution on the initial
conditions can be exploited to generate assemblies of aggregates with
tailored size distributions.
Moreover, in situations where $k$ shows a non-trivial evolution in time the
present theory provides a more natural starting point for an analysis
of the aggregate growth than the Lifshitz-Slyozov-Wagner theory,
because according to \Eq{define-k} the point $k=1$ is unstable with
respect to growth of $k$ when $n$ decreases due to the evaporation of
aggregates.

\begin{acknowledgments}

We acknowledge feedback by Karl-Henning Rehren upon developing the
present theory, and inspiring discussions with 
Markus Abel,
Bernhard Altaner, 
Nicolas Rimbert, 
Artur Wachtel, and 
Michael Wilkinson. 
Lucas Goehring, Stephan Herminghaus, and Artur Wachtel provided feedback 
on the manuscript.

\end{acknowledgments}


%

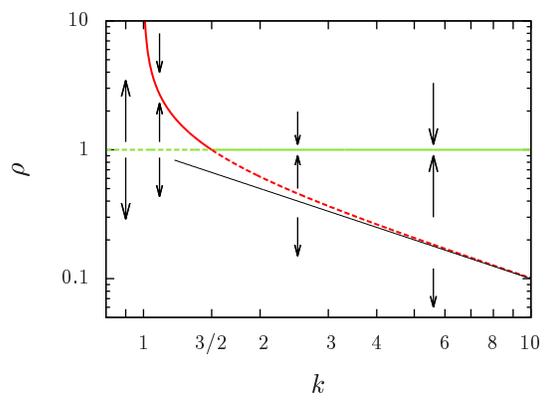
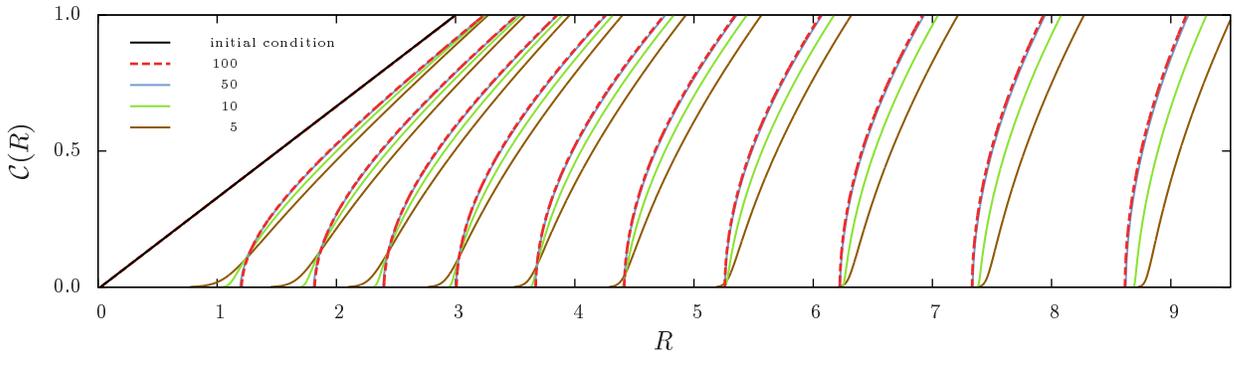
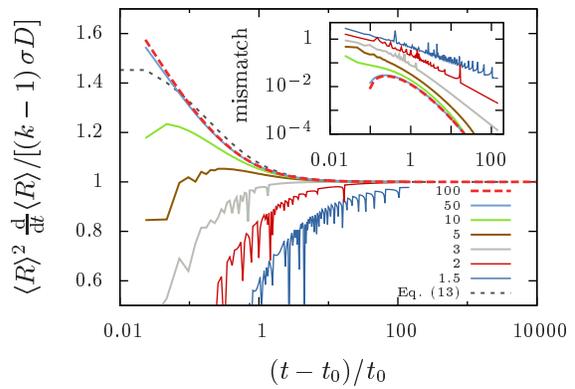
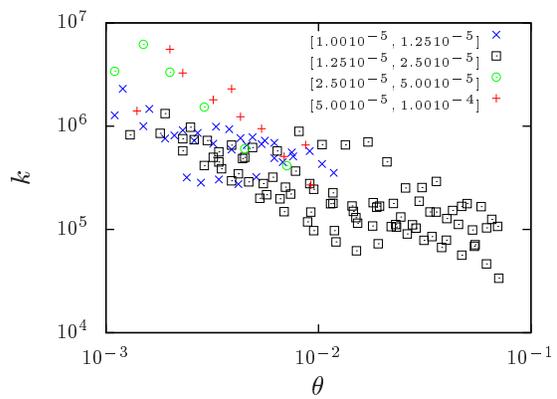
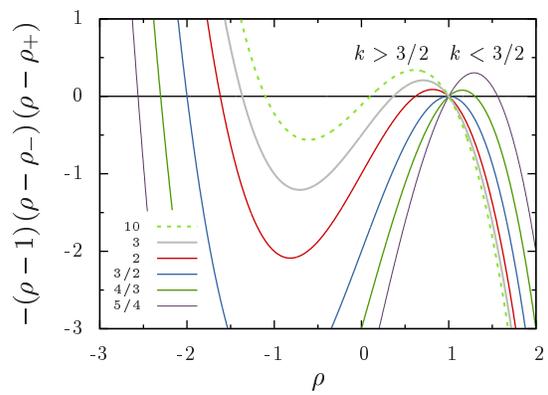
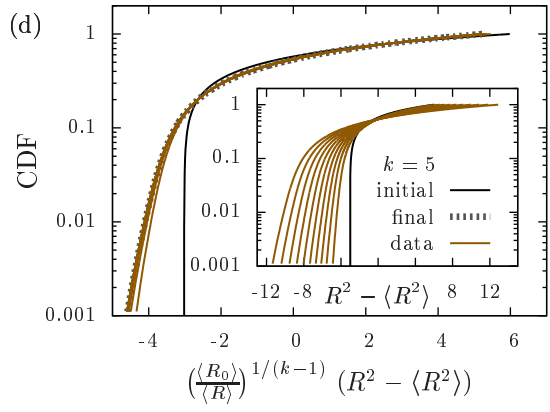
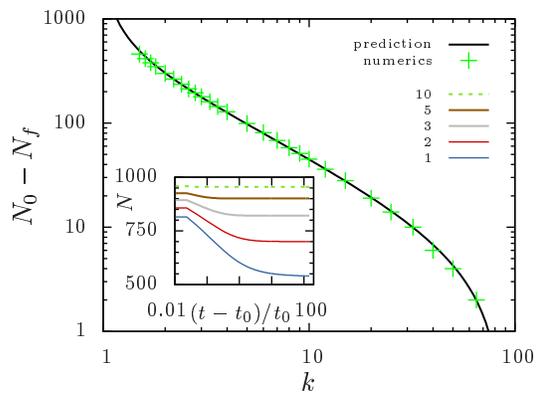
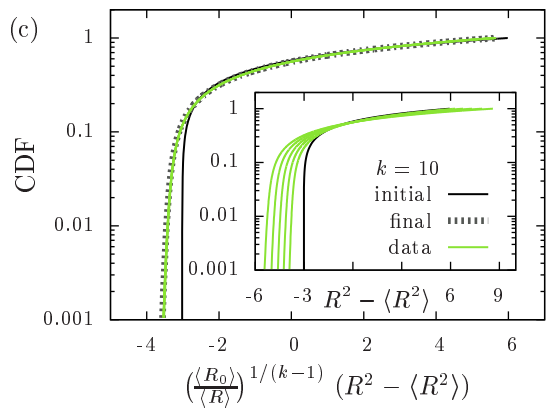
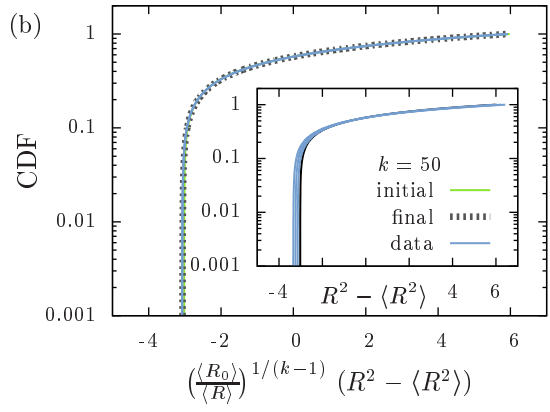
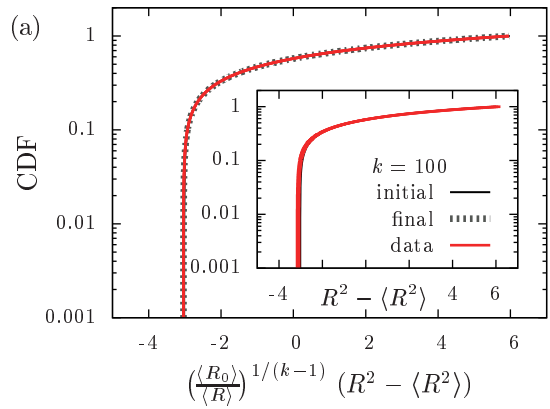
\end{document}